\documentclass[aps,pre,twocolumn]{revtex4}

\usepackage{color}
\usepackage{subfig}
\usepackage{graphicx}
\usepackage{units}
\usepackage[version=3]{mhchem}
\usepackage[percent]{overpic}

\begin{document}

\title{ Theory of Chemical Kinetics and 
Charge Transfer \\ based on Nonequilibrium Thermodynamics }
\author{Martin Z. Bazant }
\email{bazant@mit.edu}
\affiliation{Departments of Chemical Engineering and Mathematics, Massachusetts Institute of Technology, Cambridge, Massachusetts 02139, USA}


\begin{abstract}
\centerline{\bf CONSPECTUS}
Advances in the fields of catalysis and electrochemical energy conversion often involve nanoparticles, which can have kinetics surprisingly different from the bulk material. Classical theories of chemical kinetics assume independent reactions in dilute solutions, whose rates are determined by mean concentrations. In condensed matter, strong interactions alter chemical activities and create variations that can dramatically affect the reaction rate. The extreme case is that of a reaction coupled to a phase transformation, whose kinetics must depend not only on the order parameter, but also its gradients at phase boundaries. Reaction-driven phase transformations are common in electrochemistry, when charge transfer is accompanied by ion intercalation or deposition in a solid phase. Examples abound in Li-ion, metal-air, and lead-acid batteries, as well as metal electrodeposition/dissolution. In spite of complex thermodynamics, however, the standard kinetic model is the Butler-Volmer equation, based on a dilute solution approximation. The Marcus theory of charge transfer likewise considers isolated reactants and neglects elastic stress, configurational entropy, and other non-idealities in condensed phases. 

The limitations of existing theories recently became apparent for the Li-ion battery material, Li$_x$FePO$_4$ (LFP). It has a strong tendency to separate into Li-rich and Li-poor solid phases, which scientists believe limits its performance. Chemists first modeled phase separation in LFP as an isotropic ``shrinking core" within each particle, but experiments later revealed striped phase boundaries on the active crystal facet. This raised the question: What is the reaction rate at a surface undergoing a phase transformation? Meanwhile, dramatic rate enhancement was attained with LFP nanoparticles, and classical battery models could not predict the roles of phase separation and surface modiÞcation. 

In this Account, I present a general theory of chemical kinetics, developed over the past seven years, which is capable of answering these questions. The reaction rate is a nonlinear function of the thermodynamic driving force -- the free energy of reaction -- expressed in terms of variational chemical potentials. The theory unifies and extends the Cahn-Hilliard and Allen-Cahn equations through a master equation for non-equilibrium chemical thermodynamics. For electrochemistry, I have also generalized both Marcus and Butler-Volmer kinetics for concentrated solutions and ionic solids. 

This new theory provides a quantitative description of LFP phase behavior. Concentration gradients and elastic coherency strain enhance the intercalation rate. At low currents, the charge-transfer rate is focused on  exposed phase boundaries, which propagate as ``intercalation waves", nucleated by surface wetting. Unexpectedly, homogeneous reactions are favored above a critical current and below a critical size, which helps to explain the rate capability of LFP nanoparticles. Contrary to other mechanisms, elevated temperatures and currents may enhance battery performance and lifetime by suppressing phase separation. The theory has also been extended to porous electrodes and could be used for battery engineering with multiphase active materials. 

More broadly, the theory describes non-equilibrium chemical systems at mesoscopic length and time scales, beyond the reach of molecular simulations and bulk continuum models. The reaction rate is consistently defined for inhomogeneous, non-equilibrium states; for example, with phase separation, large electric fields, or mechanical stresses. This research is also potentially applicable to fluid extraction from nanoporous solids, pattern formation in electrophoretic deposition, and electrochemical dynamics in biological cells. 
 \end{abstract}

\maketitle

\begin{figure*}[tb]
\begin{center}
(a) \includegraphics[width=1.8in]{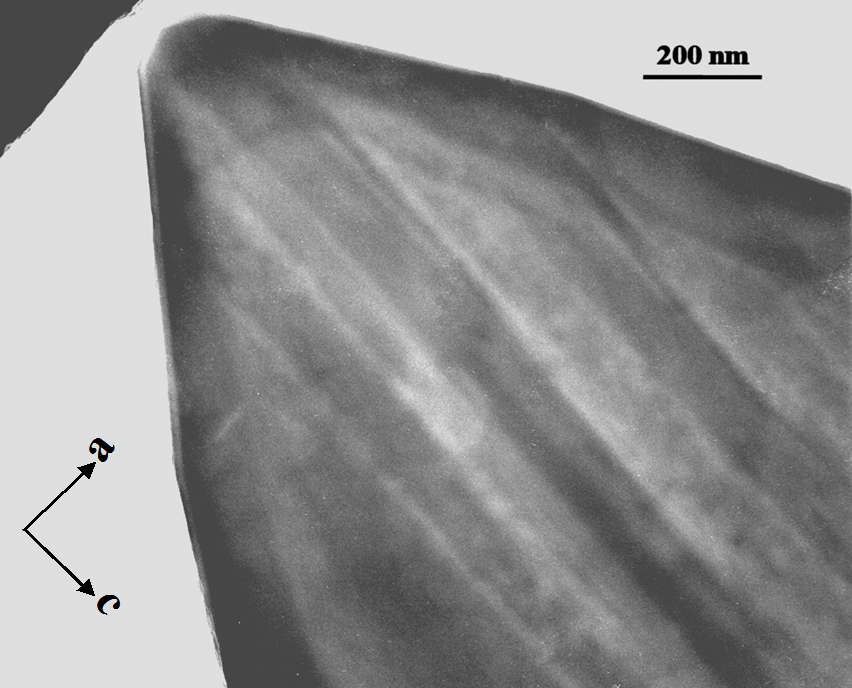}
(b) \includegraphics[width=1.5in]{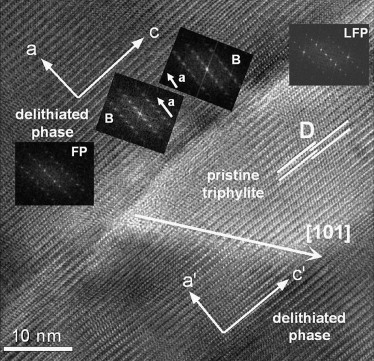}
(c)\includegraphics[width=2.4in]{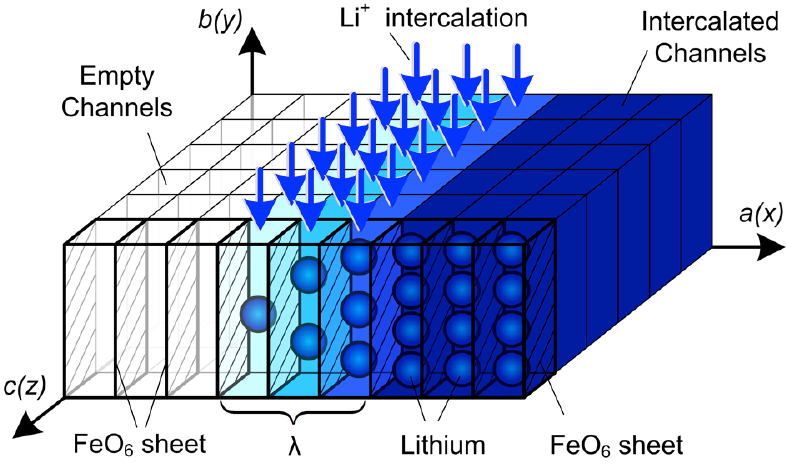}
\caption{ Motivation to generalize charge-transfer theory.  Observations  by (a) \citet{chen2006} and  (b) \citet{ramana2009} of separated FePO$_4$ and LiFePO$_4$ phases on the active \{010\} facet, which suggest (c) focusing of lithium intercalation reactions on the phase boundary, so it propagates as an ``intercalation wave"~\cite{singh2008} (or ``domino cascade"~\cite{delmas2008}).  \label{fig:wave} [From Refs. \cite{chen2006,ramana2009,singh2008}] }
\end{center}
\end{figure*}

\section{ Introduction }

Breakthroughs in catalysis and electrochemical energy conversion often involve nanoparticles, whose kinetics can differ unexpectedly from the bulk material.  Perhaps the most remarkable case is lithium iron phosphate, Li$_x$FePO$_4$ (LFP). In the seminal study of micron-sized LFP particles, Padhi et al. \cite{padhi1997} concluded that ``the material is very good for  low-power applications" but ``at
higher current densities there is a reversible decrease in capacity
that... is associated with the movement of a two-phase interface" between LiFePO$_4$ and FePO$_4$. Ironically, over the next decade -- in nanoparticle form -- LFP became the most popular  high-power cathode material for Li-ion batteries~\cite{tarascon2001,kang2009,tang2010}.  Explaining this reversal of fortune turned out to be a major scientific challenge, with important technological implications.

It is now understood that phase separation is strongly suppressed in LFP nanoparticles, to some extent in equilibrium~\cite{meethong2007,burch2009,cogswell2012,cogswell2013}, but especially under applied current~\cite{bai2011,cogswell2012,malik2011,wagemaker2011}, since reaction limitation~\cite{singh2008}, anisotropic lithium transport~\cite{morgan2004,laffont2006,delmas2008,malik2010},  elastic coherency strain~\cite{meethong2007a,vanderven2009,cogswell2012,stanton2012}, and interfacial energies~\cite{vanderven2009a,wagemaker2009,bai2011,cogswell2013} are all enhanced. At low currents, anisotropic nucleation and growth can also occur~\cite{singh2008,bai2011,oyama2012,cogswell2012,cogswell2013}, as well as  multi-particle mosaic instabilities~\cite{dreyer2010,dreyer2011,burch_thesis,ferguson2012}. These complex phenomena cannot be described by traditional battery models~\cite{doyle1993,newman_book}, which assume a spherical ``shrinking core" phase boundary~\cite{srinivasan2004,dargaville2010}.

This Account summarizes my struggle to develop a phase-field theory of electrochemical kinetics~\cite{10.626,singh2008,burch2008wave,burch2009,bai2011,cogswell2012,ferguson2012,stanton2012,cogswell2013,large_acis} by combining charge-transfer theory~\cite{kuznetsov_book} with concepts from statistical physics~\cite{sekimoto_book} and non-equilibrium thermodynamics~\cite{degroot_book,kom,prigogine_book}. 
It all began in 2006 when my postdoc, Gogi Singh, found the paper of \citet{chen2006}  revealing striped phase boundaries in LFP, looking nothing like a shrinking core and suggesting phase boundary motion perpendicular to the lithium flux (Fig. \ref{fig:wave}). It occurred to me that, at such a surface, intercalation reactions must be favored on the phase boundary in order to preserve the stable phases, but this could not be described by classical kinetics proportional to concentrations. Somehow the reaction rate had to be sensitive to concentration  gradients.

As luck would have it, I was working on models of charge relaxation in concentrated electrolytes using non-equilibrium thermodynamics~\cite{kilic2007b,large_acis}, and this seemed like a natural starting point.  Gerbrand Ceder suggested adapting the Cahn-Hilliard (CH) model for LFP~\cite{han2004}, but it took several years to achieve a consistent theory. Our initial manuscript~\cite{singh2007} was rejected in 2007, just after Gogi left MIT and I went on sabbatical leave to ESPCI, faced with rewriting the paper~\cite{singh2008}. 

The rejection was a blessing in disguise, since it made me think harder about the foundations of chemical kinetics.
The paper contained some new ideas -- phase-field chemical kinetics and intercalation waves -- that, the reviewers felt, contradicted the laws of electrochemistry.  It turns out the basic concepts were correct, but Ken Sekimoto and David Lacoste at ESPCI helped me realize that my initial Cahn-Hilliard reaction  (CHR)  model did not uphold the De Donder relation~\cite{sekimoto_book}. 
In 2008 in Paris, I completed the theory, prepared lecture notes~\cite{10.626}, published generalized Butler-Volmer kinetics~\cite{large_acis} (Sec. 5.4.2) and formulated non-equilibrium thermodynamics for porous electrodes~\cite{ferguson2012}. (See also ~\citet{sekimoto_book}.)

Phase-field kinetics represents a paradigm shift in chemical physics, which my group has successfully applied to Li-ion batteries.  Damian Burch~\cite{burch2009}  used the CHR model to study intercalation in nanoparticles, and his thesis~\cite{burch_thesis} included early simulations of ``mosaic instability" in collections of bistable particles ~\cite{dreyer2010,dreyer2011}.  Simulations of galvanostatic discharge by Peng Bai and Daniel Cogswell led to the unexpected prediction of a critical current for the suppression of phase separation~\cite{bai2011}.  Liam Stanton modeled anisotropic coherency strain~\cite{stanton2012}, which Dan added to our LFP model~\cite{cogswell2012}, along with surface wetting~\cite{cogswell2013}. Using material properties from {\it ab initio} calculations, Dan predicted phase behavior in LFP~\cite{cogswell2012} and the critical voltage for nucleation~\cite{cogswell2013} in excellent agreement with experiments.  Meanwhile, Todd Ferguson~\cite{ferguson2012} did the first simulations of phase separation in porous electrodes, paving the way for engineering applications. 

What follows is a general synthesis of the theory and a summary its key predictions. A thermodynamic framework is developed for chemical kinetics, whose application to charge transfer generalizes the classical Butler-Volmer and Marcus equations. The  theory is then unified with phase-field models and applied to Li-ion batteries.

\section{ Reactions in Concentrated Solutions }

\subsection{ Generalized Kinetics }

The theory is based on chemical thermodynamics.  In an open system, the chemical potential of species $i$ (per particle),
\begin{equation}
\mu_i   = k_BT \ln a_i  +  \mu_i^\Theta = k_B T \ln \tilde{c}_i  +  \mu_i^{ex}
\end{equation}
is defined relative to a standard state ($\Theta$) of unit activity ($a_i=1$) and concentration $c_i=c_i^\Theta$, where $\tilde{c}_i=c_i/c_i^\Theta$ is the dimensionless concentration. The activity coefficient,
 \begin{equation}
 \gamma_i = e^{(\mu_i^{ex}-\mu_i^\Theta)/k_BT}
 \end{equation}
 is a measure of non-ideality ($a_i=\gamma_i \tilde{c}_i$). In a
 dilute solution, $\mu_i^{ex}=\mu_i^{\Theta}$ and $\gamma_i=1$.  
For the general chemical reaction, 
\begin{equation}
\mbox{S}_1 = \sum_r s_r \mbox{A}_r \to  \sum_p s_p \mbox{B}_p = \mbox{S}_2,  \label{eq:genreact}
\end{equation}
the equilibrium constant is 
\begin{equation}
K^\Theta=\left(\frac{a_2}{a_1}\right)^{eq}=e^{(\mu_1^\Theta - \mu_2^\Theta)/k_BT}
\end{equation}
where $a_1 = \prod_r a_r^{s_r}$,  $a_2 = \prod_p a_p^{s_p}$, 
$\mu_1^\Theta = \sum_r s_r \mu_r^\Theta$ and $\mu_2^\Theta = \sum_p s_p \mu_p^\Theta$. 

The theory assumes that departures from equilibrium obey linear irreversible thermodynamics (LIT)~\cite{degroot_book,kom}. The flux of species $i$ is proportional to the thermodynamic driving force $-\nabla \mu_i$:
\begin{equation}
F_i = - M_i c_i \nabla \mu_i = -D_i \left( \nabla c_i + c_i \nabla \frac{\mu_i^{ex}}{k_B T} \right) 
 = - D_i^{chem} \nabla c_i
  \label{eq:flux}
\end{equation}
where $M_i$ is the mobility, $D_i = M_i k_B T$ is the tracer diffusivity, and $D_i^{chem} = D_i \left( 1+ \frac{\partial \ln \gamma_i}{\partial \ln c_i}\right)$ is the chemical diffusivity~\cite{newman_book}.  
In Eq. \ref{eq:flux}, the first term represents random fluctuations and the second, drift in response to the thermodynamic bias, $-\nabla \mu_i^{ex}$.

\begin{figure}[t]
\includegraphics[width=3in]{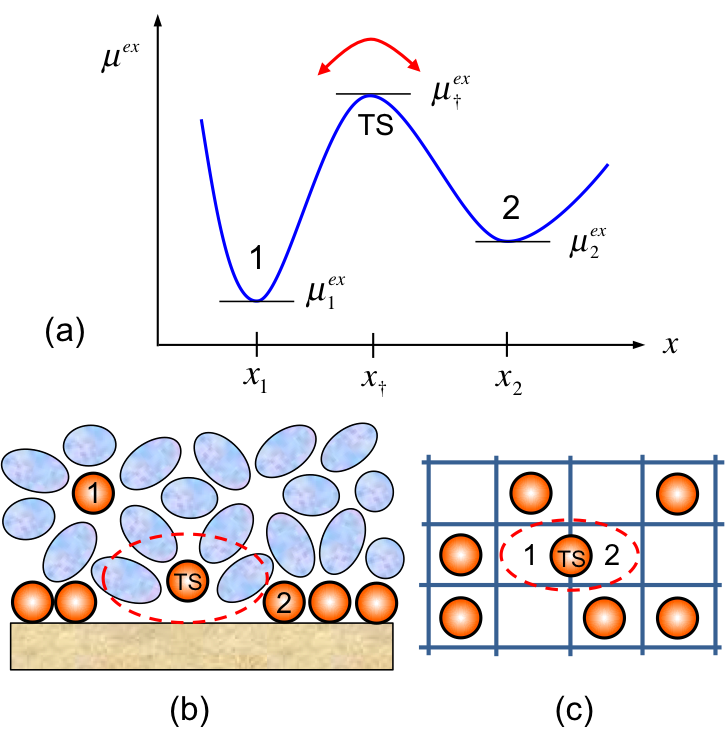}
\caption{ (a) Landscape of  excess chemical potential explored by the reaction $S_1 \to S_2$. (b) Adsorption from a liquid, where the transition state (TS) excludes multiple surface sites ($s>1$) while shedding the first-neighbor shell. (c) Solid diffusion on a lattice, where the transition state excludes two sites.
\label{fig:reaction}    }
\end{figure}

In a consistent formulation of reaction kinetics~\cite{10.626,sekimoto_book}, therefore,  the reaction complex explores  a landscape of  excess chemical potential $\mu^{ex}(x)$ between local minima $\mu_1^{ex}$ and $\mu_2^{ex}$ with transitions over an activation barrier $\mu^{ex}$  (Fig. \ref{fig:reaction}(a)).
For rare transitions ($\mu^{ex}_\ddag-\mu_{1,2}^{ex}\gg k_BT$), the reaction rate (per site) is  
\begin{equation}
R = k_\rightarrow \tilde{c}_1 e^{-(\mu^{ex}_\ddag - \mu^{ex}_1)/k_BT}
- k_\leftarrow \tilde{c}_2 e^{-(\mu^{ex}_\ddag - \mu^{ex}_2)/k_BT}
\end{equation}
Enforcing detailed balance ($R=0$) in equilibrium ($\mu_1=\mu_2$) yields the reaction rate consistent with non-equilibrium thermodynamics:
\begin{equation}
\boxed{ R = k_0 \left(e^{-(\mu^{ex}_\ddag - \mu_1)/k_BT} - e^{-(\mu^{ex}_\ddag - \mu_2)/k_BT}\right)   }
\label{eq:Rgen} \\
\end{equation}
where $k_0 = k_\rightarrow = k_\leftarrow$ (for properly defined $\mu$). Eq. \ref{eq:Rgen} upholds the De Donder relation~\cite{sekimoto_book},
\begin{equation}
\frac{R_\rightarrow}{R_\leftarrow} = \frac{ K^\Theta a_1}{a_2} = e^{(\mu_1 - \mu_2)/k_BT}
\end{equation}
which describes the steady state of chemical reactions in open systems~\cite{beard2007}.

The thermodynamic driving force is 
\begin{equation}
\Delta \mu = \mu_2 - \mu_1= k_BT \ln \frac{a_2}{K^\Theta a_1} = \Delta G
\end{equation}
also denoted as $\Delta G$, the free energy of reaction. The reaction rate Eq. \ref{eq:Rgen} can be expressed as a  nonlinear function of $\Delta \mu$:
\begin{equation}
R = R_0 \left( e^{-\alpha \Delta\mu/k_BT} - e^{(1-\alpha) \Delta\mu/k_BT} \right)  \label{eq:RBV}
\end{equation}
where $\alpha$, the symmetry factor or generalized Br{\o}nsted coefficient~\cite{kuznetsov_book}, is approximately constant with $0 < \alpha < 1$ for many reactions. Defining the activity coefficient of the transition state $\gamma_\ddag$ by
\begin{equation}
\mu^{ex}_\ddag = k_BT \ln \gamma_\ddag + (1-\alpha)\mu_1^\Theta + \alpha \mu_2^\Theta,
\end{equation}
the exchange rate $R_0$ takes the form,
\begin{equation}
R_0 = \frac{ k_0 a_1^{1-\alpha} a_2^\alpha }{\gamma_\ddag} = k_0 \tilde{c}_1^{1-\alpha} \tilde{c}_2^{\alpha}
\left( \frac{ \gamma_1^{1-\alpha} \gamma_2^\alpha}{\gamma_\ddag} \right)   \label{eq:R_0}
\end{equation}
where the term in parentheses is the thermodynamic correction for a concentrated solution.

\subsection{ Example: Surface Adsorption }

Let us apply the formalism to Langmuir adsorption, $A\to A_{ads}$, from a liquid mixture 
with $\mu_1 = k_BT \ln a$ (Fig. \ref{fig:reaction}(b)). The surface is an ideal solution of adatoms and vacancies, 
\begin{equation}
\mu_2 = k_BT \ln \frac{\tilde{c}}{1-\tilde{c}} + E_{ads}
\end{equation}
with coverage $\tilde{c}=c/c_s$, site density $c_s$, and adsorption energy $E_{ads} = \mu_2^\Theta - \mu_1^\Theta$. Equilibrium yields the Langmuir isotherm,
\begin{equation}
\tilde{c}_{eq} = \frac{K^\Theta a}{1 + K^\Theta a}, \ \ \ K^\Theta = e^{-E_{ads}/k_BT}
\end{equation}
If the transition state excludes $s$ surface sites,
\begin{equation}
\mu^{ex}_\ddag = - s k_B T \ln(1-\tilde{c}) + E_\ddag   \label{eq:gamlang}
\end{equation}
then Eq. \ref{eq:Rgen} yields,
\begin{equation}
R =  k_1 (1-\tilde{c})^s \left[ K^\Theta a (1-\tilde{c}) - \tilde{c} \right]
\end{equation}
where $k_1=k_0 e^{(E_{ads}-E_\ddag)/k_BT}$ and $E_{\ddag}$ is the transition state energy relative to the bulk.
With only configurational entropy, we recover standard kinetics of adsorption,  $A_{sol} + s V \to A_{surf} + (s-1)V$, involving $s$ vacancies. With attractive forces, however, Eq. \ref{eq:Rgen} predicts novel kinetics for inhomogeneous surfaces undergoing condensation (below).

\subsection{ Example: Solid diffusion} 
We can also derive the LIT flux Eq. \ref{eq:flux} for macroscopic transport in a solid by activated hopping between adjacent minima of $\mu^{ex}$ having slowly varying chemical potential, $|\Delta \mu| \ll k_BT$ and concentration $\Delta \tilde{c} \ll 1$. Linearizing the hopping rate, 
\begin{equation}
R \sim -\frac{R_0\Delta \mu}{k_BT}, \ \  R_0 \sim \frac{ k_0 \tilde{c} \gamma}{\gamma_\ddag}
\end{equation}
over a distance $\Delta x$ through an area $\Delta y \Delta z$ with $\frac{\partial \mu}{\partial x} \sim \frac{\Delta\mu}{\Delta x}$, we obtain Eq. \ref{eq:flux} with 
\begin{equation}
\frac{D}{D_0} = \frac{\gamma}{\gamma_\ddag}   \label{eq:D}
\end{equation}
where $D_0 = \frac{k_0 \Delta x}{c^\Theta \Delta y \Delta z}$.
Eq. \ref{eq:D} can be used to derive the tracer diffusivity in a concentrated solid solution by estimating $\gamma_\ddag$, consistent with $\gamma$. For example,  for diffusion on a lattice (Fig. \ref{fig:reaction}(c)) with $\gamma = (1-\tilde{c})^{-1}$, the transition state excludes two sites, $\gamma_\ddag = (1-\tilde{c})^{-2}$; the tracer diffusivity, $D=D_0 (1-\tilde{c})$, scales with the mean number of empty neighboring sites, but  the chemical diffusivity is constant, $D^{chem} = D_0 = D(0)$ (particle/hole duality).

\begin{figure}[t]
\includegraphics[width=2.5in]{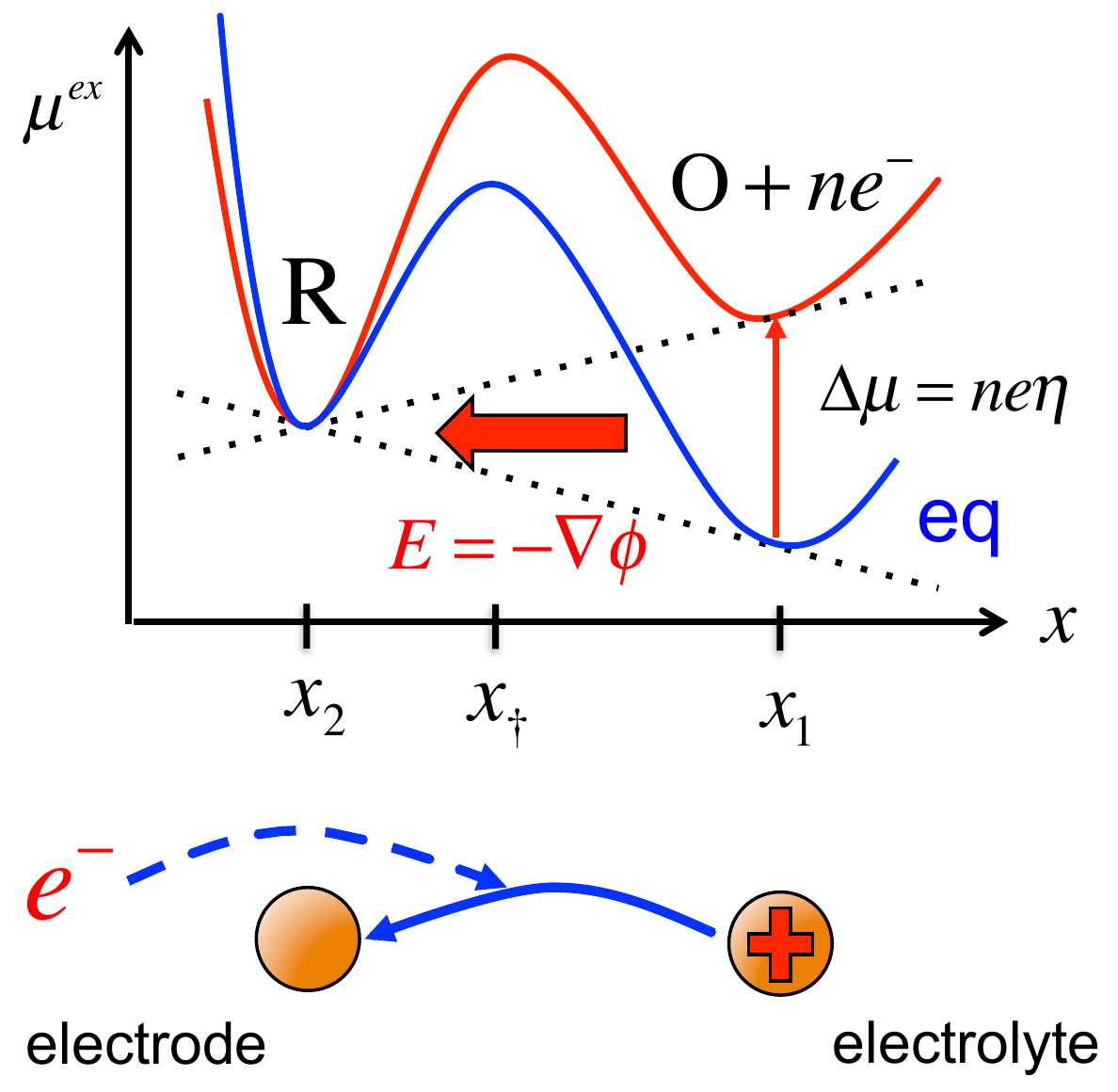}
\caption{ Landscape of excess chemical potential explored by the Faradaic reaction $\mbox{O} + n e^- \to \mbox{R}$, in Nernst equilibrium (blue) and after a negative overpotential $\eta = (\mu_2-\mu_1)/ne$ is applied (red) to favor reduction, as illustrated below. \label{fig:BV}    }
\end{figure}

\section{ Electrochemistry in Concentrated Solutions }

\subsection{ Electrochemical Thermodynamics }

Next we apply Eq. \ref{eq:Rgen} to the general Faradaic reaction, 
\begin{equation}
S_1 =  \sum_i s_{i,O} \mbox{O}_i^{z_{i,O}} + n e^-  \rightarrow 
\sum_j s_{j,R} \mbox{R}_j^{z_{j,R}} = S_2   \label{eq:Faradaic}
\end{equation}
converting the oxidized state $\mbox{O}^{z_O}= \sum_i s_{i,O} \mbox{O}_i^{z_{i,O}}$ to the reduced state $\mbox{R}^{z_R}=\sum_j s_{j,R} \mbox{R}_j^{z_{j,R}}$ while consuming $n$ electrons. Let  $\mu_1 = \mu_O + n \mu_e = \sum_i s_{i,O} \mu_{i,O} + n \mu_e$ and  $\mu_2 = \mu_R = \sum_j s_{j,r} \mu_{j,r}$. Charge conservation implies $z_O - n = z_R$ where $z_O=\sum_i s_{i,O} z_{i,O}$ and $z_R=\sum_j s_{j,R} z_{j,R}$.  The electrostatic energy $z_i e \phi_i$ is added to $\mu_i^{ex}$ to define the electrochemical potential,
\begin{equation}
\mu_i = k_BT \ln a_i  + \mu_i^\Theta + z_i e \phi_i = k_BT \ln \tilde{c}_i + \mu_i^{ex} 
\end{equation}
where $z_i e$ is the charge and $\phi_i$ is the Coulomb potential of mean force.   

\begin{figure}[tb]
\includegraphics[width=2.5in]{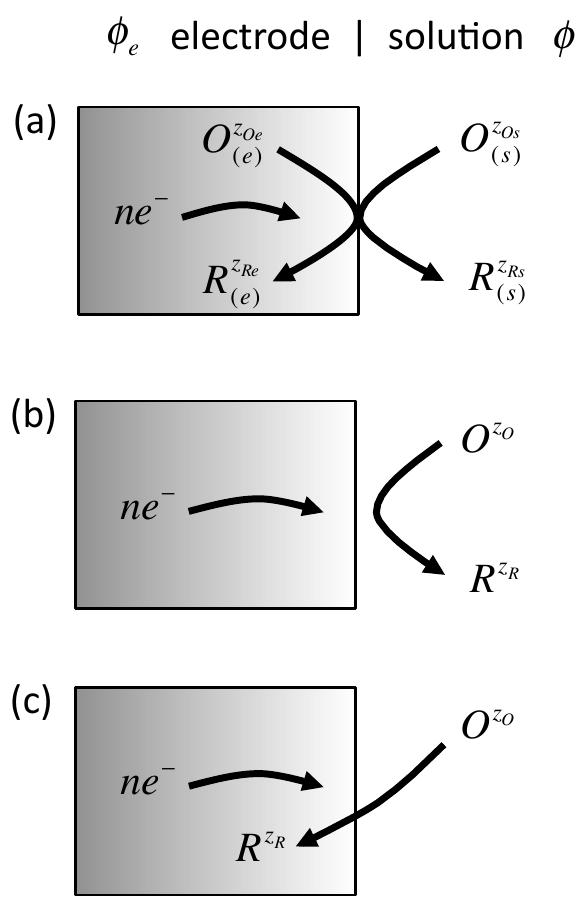}
\caption{  Types of Faradaic reactions $\mbox{O} + ne^- \to \mbox{R}$. (a) General mixed ion-electron conductor electrode/electrolyte interface. (b) Redox in solution. (c) Ion intercalation or electrodeposition. \label{fig:Faradaic}    }
\end{figure}

The electrostatic potential is $\phi_e$ in the electrode and $\phi$ in the electrolyte.  The difference is the interfacial voltage, $\Delta\phi = \phi_e - \phi$. The mean electric field $-\nabla\phi$ at a point is unique, so $\phi_i=\phi_e$ for ions in the electrode and $\phi_i=\phi$ for those in the electrolyte solution. In the most general case of a mixed ion-electron conductor, the reduced and oxidized states are split across the interface (Fig. \ref{fig:Faradaic}(a)). Charge conservation implies  
$z_{Oe}+z_{Os}-n=z_{Re}+z_{Rs}$, and the net charge $n_c e$ transferred from the solution to the electrode is given by $n_c = z_{Os} - z_{Rs} = z_{Re}-z_{Oe}+n$.  

Let us assume that ions only exist in the electrolyte ($z_{Re}=z_{Oe}=0,n_c=n$) since the extension to mixed ion-electron conductors is straightforward. For redox reactions (Fig. \ref{fig:Faradaic}(b)),  e.g.  Fe$^{3+} + e^-\to$ Fe$^{2+}$, the reduced state is in the solution at the same potential, $\phi_R = \phi_O = \phi$.  For electrodeposition (Fig. \ref{fig:Faradaic}(c)), e.g. Cu$^{2+} + 2 e^- \to $ Cu,  or ion intercalation as a neutral polaron, e.g. CoO$_2 +$Li$^+ + e^- \to $ LiCoO$_2$, the reduced state is uncharged, $z_R=0$, so we can also set $\phi_R=\phi$, even though it is in the electrode. For this broad class of Faradaic reactions,  we have 
\begin{equation}
\mu_O = k_BT \ln a_O + \mu_O^\Theta + z_O e \phi   \label{eq:muodef} 
\end{equation}
\begin{equation}
\mu_R = k_BT \ln a_R + \mu_R^\Theta + z_R e \phi 
\end{equation}
\begin{equation}
\mu_e = k_BT \ln a_e + \mu_e^\Theta - e  \phi_e   \label{eq:muedef}
\end{equation} 
($a_O=\prod_i a_i^{s_j}$, $\mu_O^\Theta = \sum_i s_i \mu_i^{\Theta}$,...) where $\mu_e$ is the Fermi level, which depends on $\phi_e$ and the electron activity $a_e=\gamma_e \tilde{c}_e$.

In equilibrium ($\mu_1 = \mu_2$), the interfacial voltage is given by the Nernst equation
\begin{equation}
\Delta\phi^{eq} = E^\Theta+ \frac{k_B T}{n_c e} \ln \frac{a_O a_e^n}{a_R} \label{eq:Nernst}
\end{equation}
where $n_c=n$ and 
\begin{equation}
E^\Theta = \frac{\mu_O^\Theta + n \mu_e^\Theta - \mu_R^\Theta}{ne}
\end{equation}
is the standard half-cell potential.
Out of equilibrium, the current $I = n e R$ (per active site) is controlled by the activation over-potential, 
\begin{equation}
\eta = \Delta\phi -\Delta \phi^{eq} = \frac{\Delta \mu}{n e} = \frac{\Delta G}{ne}  \label{eq:eta}
\end{equation}
Specific models of charge transfer correspond to different choices of $\mu^{ex}_\ddag$.

\subsection{ Generalized Butler-Volmer Kinetics }

The standard phenomenological model of electrode kinetics is the Butler-Volmer equation\cite{bard_book,newman_book},
\begin{equation}
I = I_0 \left(e^{-\alpha_c ne \eta/k_BT} - e^{\alpha_a ne \eta/k_BT}\right) 
\label{eq:BV}
\end{equation}
where $I_0$ is the exchange current $I_0$.
For a  single-step charge-transfer reaction, the anodic and cathodic charge-transfer coefficients $\alpha_a$ and $\alpha_c$ satisfy $\alpha_a=1-\alpha$ and $\alpha_c=\alpha$ with a symmetry factor, $0<\alpha<1$. The exchange current is typically modeled as 
$I_0 \propto c_O^{\alpha_a} c_R^{\alpha_c}$,   
but this is a  dilute solution approximation. 

In concentrated solutions, the exchange current is affected by configurational entropy and enthalpy, electrostatic correlations, coherency strain, and other non-idealities.  For Li-ion batteries, only excluded volume has been considered, using\cite{doyle1993,newman_book}, $I_0(c) \propto (c_s-c)^{\alpha_c} c^{\alpha_a}$.
For fuel cells, many phenomenological models have been developed for electrocatalytic reactions with surface adsorption steps~\cite{kulikovsky_book,eikerling1998,ioselevich2001}.  Electrocatalysis can also be treated by our formalism~\cite{10.626}, but here we focus on the elementary charge-transfer step and its coupling to phase transformations, which has no prior literature.  

In order to generalize BV kinetics (Fig. \ref{fig:BV}), we model the transition state 
\begin{equation}
\mu^{ex}_\ddag = k_B T \ln \gamma_\ddag + (1-\alpha) (z_O e\phi - ne\phi_e + \mu_O^\Theta + n \mu_e^\Theta) + \alpha (z_R e \phi + \mu_R^\Theta)   \label{eq:bvts}
\end{equation}
by averaging the standard chemical potential and electrostatic energy of the initial and final states, which assumes a constant electric field across the reaction coordinate $x$ with $\alpha = \frac{x_\ddag-x_R}{x_O-x_R}$. Substituting Eq. \ref{eq:bvts} into Eq. \ref{eq:Rgen} using Eq. \ref{eq:Nernst}, we obtain  Eq. \ref{eq:BV} with
\begin{equation}
\boxed{ I_0  = 
 \frac{k_0 ne (a_O a_e^n)^{1-\alpha}a_R^\alpha}{\gamma_\ddag} }
= 
k_0 ne (\tilde{c}_O \tilde{c}_e^n)^{1-\alpha}\tilde{c}_R^\alpha
 \left[ \frac{(\gamma_O\gamma_e^n)^{1-\alpha}\gamma_R^\alpha}{\gamma_\ddag} \right]    \label{eq:I0} 
\end{equation}
The factor in brackets is the thermodynamic correction for the exchange current.  

Generalized BV kinetics (Eq. \ref{eq:BV} and  Eq. \ref{eq:I0}) consistently applies chemical kinetics in concentrated solutions (Eq. \ref{eq:RBV} and Eq. \ref{eq:R_0}, respectively) to Faradaic reactions.
 In Li-ion battery models,  $\Delta\phi^{eq}(c)$ is fitted to the open circuit voltage, and $I_0(c)$ and  
$D^{chem}(c)$ are fitted to discharge curves~\cite{doyle1993,srinivasan2004,dargaville2010}, but these quantities are related by non-equilibrium thermodynamics~\cite{large_acis,bai2011,ferguson2012}.   Lai and Ciucci~\cite{lai2010,lai2011a,lai2011b} also recognized this inconsistency and used Eq. \ref{eq:flux} and Eq. \ref{eq:Nernst} in battery models, but they postulated a barrier of total (not excess) chemical potential, in contrast to Eq. \ref{eq:Rgen}, Eq. \ref{eq:I0} and charge-transfer theory.

\begin{figure}
\includegraphics[width=2.4in]{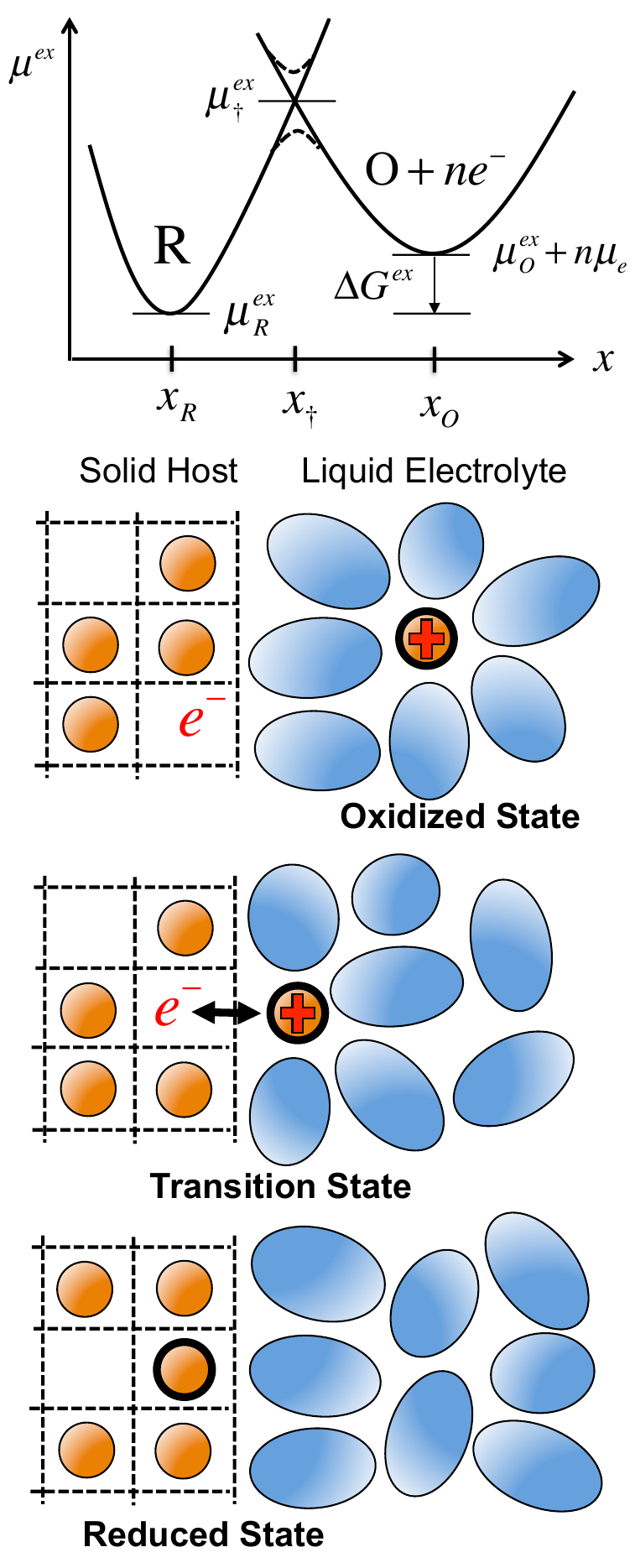}
\caption{ Above:   The Faradaic reaction $O+ne^- \to R$ in concentrated solutions. Each state explores a landscape of  excess chemical potential $\mu^{ex}$. Charge transfer occurs where the curves overlap, or just below, by quantum tunneling (dashed curves).  Below: Example of ion intercalation into a solid electrode from a liquid electrolyte.  \label{fig:Marcus}    }
\end{figure}

\subsection{ Generalized Marcus Kinetics }

The microscopic theory of charge transfer, initiated by Marcus~\cite{marcus1956,marcus1965} and honored by the Nobel Prize in Chemistry~\cite{marcus1993}, provides justification for the BV equation and a means to estimate its parameters based on solvent reorganization~\cite{bard_book}. Quantum mechanical formulations pioneered by Levich, Dogonadze, Marcus, Kuznetsov, and Ulstrup further account for Fermi statistics, band structure, and electron tunneling~\cite{kuznetsov_book}.  Most theories, however, make the dilute solution approximation by considering an isolated reaction complex. 

In order to extend Marcus theory for concentrated solutions, our basic postulate  (Fig. \ref{fig:Marcus}) is that the Faradaic reaction Eq. \ref{eq:Faradaic} occurs when the  excess chemical potential of the reduced state, deformed along the reaction coordinate by statistical fluctuations, equals that of the oxidized state  (plus $n$ electrons in the electrode) at the same point.  (More precisely, charge transfer occurs at slightly lower energies due to quantum tunneling~\cite{kuznetsov_book,bard_book}.) Following Marcus, we assume harmonic restoring forces for structural relaxation (e.g. shedding of the solvation shell from a liquid, or ion extraction from a solid) along the reaction coordinate  $x$ from the oxidized state at $x_O$ to the reduced state at $x_R$:
\begin{equation}
\mu_1^{ex}(x) =\mu_O^\Theta + n\mu_e^\Theta + k_BT \ln (\gamma_O\gamma_e^n) + z_O e \phi - ne\phi_e + \frac{k_O}{2}(x-x_O)^2 
\end{equation}
\begin{equation}
\mu_2^{ex}(x) = \mu_R^\Theta+ k_BT \ln \gamma_R + z_R e \phi + \frac{k_R}{2}(x-x_R)^2 
\end{equation}
The Nernst equation Eq. \ref{eq:Nernst} follows by equating the  total chemical potentials at the local minima, $\mu_1(x_O) = 
\mu_2(x_R)$ in equilibrium.  The free energy barrier is set by the intersection of the  excess chemical potential curves, 
$\mu^{ex}_\ddag = \mu_1^{ex}(x_\ddag) = \mu_2^{ex}(x_\ddag)$,  
which determines the barrier position, $x=x_\ddag$ and implies
\begin{equation}
\Delta G^{ex} = \mu^{ex}_2(x_R) - \mu^{ex}_1(x_O)   =   
 \frac{k_O}{2}(x_\ddag - x_O)^2 - \frac{k_R}{2}(x_\ddag - x_R)^2   \label{eq:G0}
\end{equation}
where  $\Delta G^{ex}$  is the  excess free energy change per reaction.  

From  Eq. \ref{eq:eta}, the overpotential is the  total free energy change per charge transferred, 
\begin{equation}
ne \eta = \Delta G = \Delta G^{ex} +  k_BT \ln\frac{\tilde{c}_R}{\tilde{c}_O\tilde{c}_e^n}   \label{eq:etadef}
\end{equation}
In classical Marcus theory~\cite{bard_book,marcus1993}, the overpotential is defined by  $ne\eta=\Delta G^{ex}$ without the concentration factors required by non-equilibrium thermodynamics, which is valid for charge-transfer reactions in bulk phases ($A^- + B \to A + B^-$) because the initial and final concentrations are the same, and thus $\Delta G = \Delta G^{ex} = \Delta G^0$ (standard free energy of reaction). For Faradaic reactions at interfaces, however, the  concentrations of reactions and products are different, and  Eq. \ref{eq:etadef} must be used.  The missing ``Nernst concentration term" in Eq. \ref{eq:etadef} has also been noted by ~\citet{kuznetsov_book} (p. 219).

In order to relate $\mu_\ddag^{ex}$ to $\Delta G^{ex}$, we solve Eq. \ref{eq:G0}  for $x_\ddag$. In the simplest approximation, $k_O = k_R = k$, the barriers for the cathodic and anodic reactions,
\begin{equation}
\Delta G^{ex}_c = \mu_\ddag^{ex} - \mu_1^{ex}(x_O) = 
\frac{\lambda}{4}\left(1+\frac{\Delta G^{ex}}{\lambda}\right)^2 
\end{equation}
\begin{equation}
\Delta G^{ex}_a = \mu_\ddag^{ex} - \mu_2^{ex}(x_R) 
= \frac{\lambda}{4}\left(1-\frac{\Delta G^{ex}}{\lambda}\right)^2 
\end{equation}
are related to the reorganization energy,
$\lambda = \frac{k}{2}(x_O - x_R)^2$.  These formulae contain the famous ``inverted region" predicted by Marcus for isotopic exchange~\cite{marcus1993}, where (say) the cathodic rate, $k_c \propto e^{-\Delta G^{ex}_c/k_BT}$ reaches a minimum and increases again with decreasing driving force $\Delta G^{ex}$, for $x_\ddag < x_R$ in Fig. \ref{fig:Marcus}(a). This effect remains for charge transfer in concentrated bulk solutions, e.g. $\mbox{A}^- + \mbox{B} \rightarrow \mbox{A}+ \mbox{B}^-$. For Fardaic reactions, however, it is suppressed at metal electrodes, since electrons can tunnel through unoccupied conduction-band states, but can arise in narrow-band semiconductors~\cite{marcus1965,marcus1993,kuznetsov_book}. 

Substituting $\mu_\ddag^{ex}$ into Eq. \ref{eq:Rgen}, we obtain 
\begin{eqnarray}
R = k_0 e^{-\lambda/4k_BT} e^{-(\Delta G^{ex})^2/4k_BT\lambda} \nonumber \\
\times \left(\tilde{c}_o\tilde{c}_e^n e^{-\Delta G^{ex}/2k_BT} - \tilde{c}_R e^{\Delta G^{ex}/2k_BT}\right)
\end{eqnarray}
Using Eq. \ref{eq:etadef}, we can relate the current to the overpotential,
\begin{equation}
I = I_0 \, e^{-(ne\eta)^2/4k_BT\lambda}\left( e^{-\alpha ne\eta/k_BT} - e^{(1-\alpha) ne\eta/k_BT} \right)    \label{eq:marcus}
\end{equation} 
via the exchange current,
\begin{equation}
I_0 = ne k_0 e^{-\lambda/4k_BT} (\tilde{c}_O \tilde{c}_e^n)^{\frac{3-2\alpha}{4}} 
\tilde{c}_R^{\frac{1+2\alpha}{4}},  \label{eq:marcI0}
\end{equation} 
and symmetry factor,
\begin{equation}
\alpha = \frac{1}{2}\left(1 + \frac{k_BT}{\lambda} \ln \frac{ \tilde{c}_O \tilde{c}_e^n}{\tilde{c}_R}  \right).   \label{eq:alpha}
\end{equation}
In the typical case $\lambda \gg k_BT$, the current Eq. \ref{eq:marcus} is well approximated by the BV equation with $\alpha=\frac{1}{2}$ at moderate overpotentials, $|\eta| > \frac{k_BT}{ne} \sqrt{\frac{\lambda}{k_BT}}$ and non-depleted concentrations,  $|\ln\tilde{c}|\ll \frac{\lambda}{k_BT}$. 

Comparing Eq. \ref{eq:marcI0} with Eq. \ref{eq:I0} for $\alpha \approx \frac{1}{2}$, we can relate the reorganization energy to the activity coefficients defined above,
\begin{equation}
\lambda \approx  4 k_B T \ln \frac{\gamma_\ddag } { (\gamma_O\gamma_e^n\gamma_R)^{1/2}}  \label{eq:lamgen}
\end{equation}
For a dilute solution, the reorganization energy $\lambda_0$ can be estimated by the classical Marcus approximation,  $\lambda_0 = \lambda_{i} + \lambda_{o}$, where  $\lambda_i$
is the ``inner" or short-range contribution from structural relaxation (sum over normal modes) and $\lambda_o$ is the ``outer", long-range contribution from the Born energy of solvent dielectric relaxation~\cite{marcus1993,bard_book}. 
For polar solvents at room temperature, the large Born energy, $\lambda_o > 0.5 n^2  \mbox{eV} \approx 20 n^2  k_BT$ (at room temperature), implies that single-electron ($n=1$), symmetric ($\alpha\approx\frac{1}{2}$) charge transfer is favored.  Quantum mechanical approximations of $\lambda_0$ are also available~\cite{kuznetsov_book}.
For a concentrated solution, we can estimate the thermodynamic correction, $\gamma_{\ddag}^{c}$, for the entropy and enthalpy of the transition state and write 
\begin{equation}
\gamma_\ddag = \gamma_\ddag^{c} e^{\lambda_0/4k_BT}.    \label{eq:gamgen}
\end{equation}
which can be used in either Marcus (Eqs. \ref{eq:marcus}-\ref{eq:lamgen}) or BV (Eqs. \ref{eq:BV}-\ref{eq:I0}) kinetics.
An example for ion intercalation is given below, Eq. \ref{eq:gamins}, but first we need to develop a modeling framework for chemical potentials.

\section{ Nonequilibrium Chemical Thermodynamics  }

\subsection{ General theory }   
In homogeneous bulk phases, activity coefficients depend on concentrations, but for reactions at an interface, concentration gradients must also play a role (Fig. \ref{fig:wave}). The main contribution of this work has been to formulate chemical kinetics for  inhomogeneous, non-equilibrium systems.  The most general theory appears here for the first time, building on my lectures notes~\cite{10.626}.

The theory is based the Gibbs free energy functional 
\begin{equation}
G[\{ c_i \}] = \int_V g \,dV + \oint_A \gamma_s\,dA = G_{bulk} + G_{surf}
\end{equation}
with integrals over the bulk volume $V$ and surface area $A$. 
The variational derivative~\cite{gelfand_book},
\begin{equation}
\frac{ \delta G}{\delta c_i}(x) = \lim_{\epsilon\to 0} \frac{ G[c_i(y) + \epsilon \delta_\epsilon(y-x)] - G[c_i(y)] }{\epsilon} 
\end{equation}
is the change in $G$ to add a ``continuum particle" $\delta(y-x)$ of species $i$ at point $x$, where $\delta_\epsilon(z)\to\delta(z)$ is a finite-size approximation for a particle that converges weakly (in the sense of distributions) to the Dirac delta function, for example, $\delta_\epsilon(z) = \frac{e^{-z^2/2\epsilon}}{\sqrt{2\pi \epsilon}}$. This is the consistent definition of diffusional chemical potential~\cite{cahn1958,kom},
\begin{equation}
\boxed{ \mu_i  = \frac{ \delta G}{\delta c_i} }   
\label{eq:muvar}
\end{equation}
If $g$ depends on $\{ c_i \}$ and  $\{ \nabla c_i \}$,  then
\begin{equation}
\mu_i = \frac{\partial g}{\partial c_i} - \nabla\cdot \frac{\partial g}{\partial \nabla c_i}
\end{equation}
The continuity of $\mu_i$ at the surface yields the ``natural boundary condition",
\begin{equation}
\hat{n}\cdot\frac{\partial g}{\partial \nabla c_i} = \frac{\partial \gamma_s }{\partial c_i}.   \label{eq:wet}
\end{equation}
We can also express the activity variationally, 
\begin{equation}
\boxed{ 
a_i = \exp\left(\frac{1}{k_BT} \frac{\delta G_{mix}}{\delta c_i} \right)    \label{eq:actphase}
}
\end{equation}
in terms of the free energy of mixing
\begin{equation}
G_{mix} = G_{bulk} - \sum_i \mu_i^{\Theta} \int_V c_i \, dV \label{eq:Gmix} 
\end{equation}
which we define relative to the standard states of each species.

The simplest approximation for an inhomogeneous system is the Cahn-Hilliard~\cite{cahn1958} (or Landau-Ginzburg, or Van der Waals~\cite{vdw1893}) gradient expansion, 
\begin{equation}
g = \bar{g}(\{ c_i \}) + \sum_i \left(\mu_i^\Theta c_i + \frac{1}{2} \sum_j \nabla \tilde{c}_i  \cdot \kappa_{ij} \nabla \tilde{c}_j \right)
\end{equation}
for which
\begin{equation}
\mu_i - \mu_i^\Theta = k_BT \ln a_i  = \frac{\partial \bar{g}}{\partial c_i}  - \sum_j \nabla \cdot \kappa_{ij} \nabla \frac{\tilde{c}_j}{c_j^\Theta}
\end{equation}
where $\bar{g}$ is the homogeneous free energy of mixing and
$\kappa_{ij}$ is a 2nd rank 
anisotropic tensor penalizing gradients in
components $i$ and $j$. (Higher-order derivative terms can also be added~\cite{nauman2001,bazant2011}.)

\begin{figure}[t]
\includegraphics[width=3in]{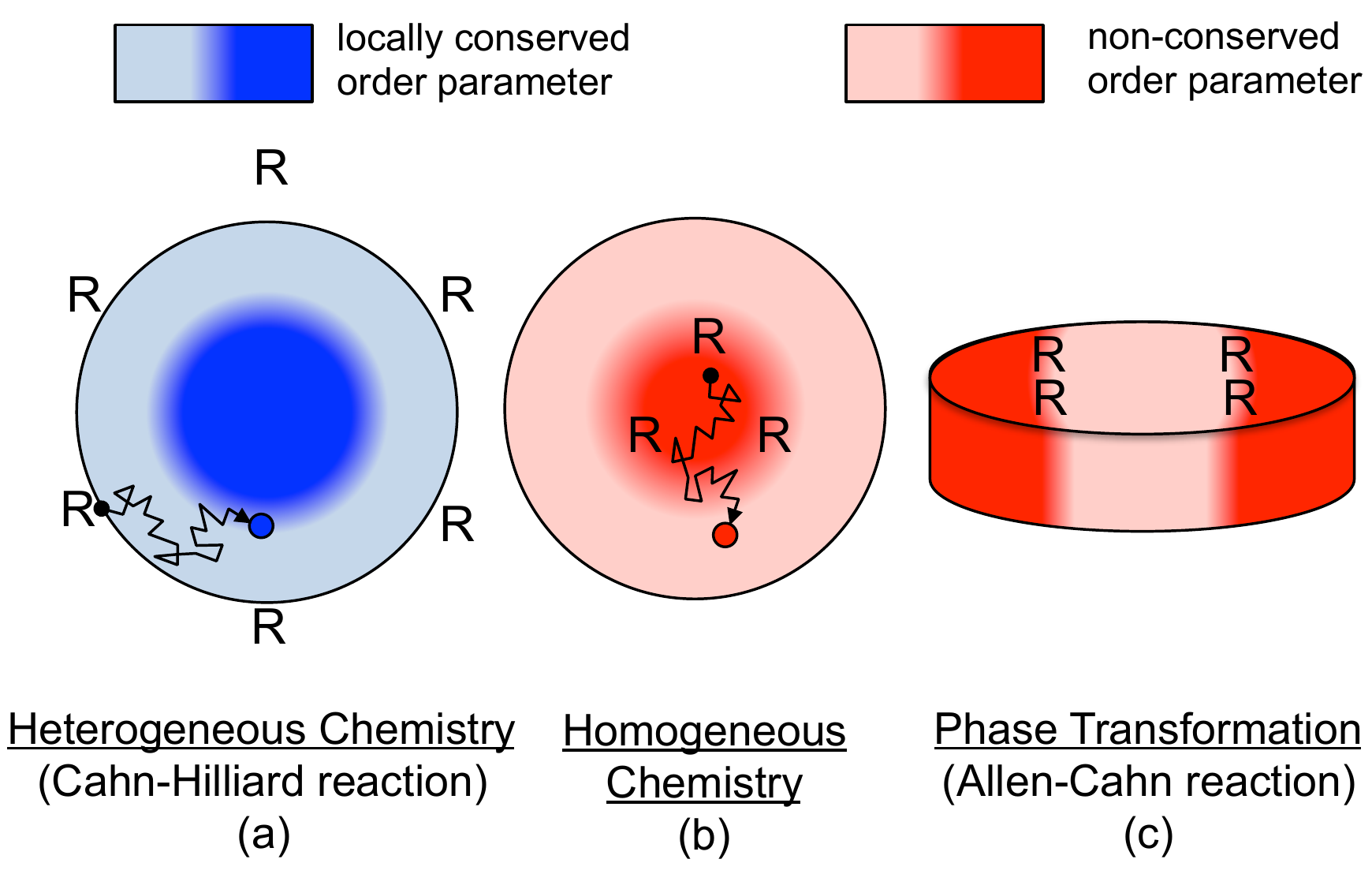}
\caption{ Types of reactions (R) in non-equilibrium chemical thermodynamics. (a) Heterogeneous chemistry at a surface Eq. \ref{eq:CHbc}. (b) Homogeneous chemistry Eq. \ref{eq:CHhomo} with diffusing species. (c) Phase transformations, or homogeneous reactions with immobile species Eq. \ref{eq:genAC}.
  \label{fig:CHR}    }
\end{figure}

With these definitions,  Eq. \ref{eq:Rgen} takes the variational form,
\begin{equation}
\boxed{ R = k_0 e^{\frac{-\mu^{ex}_\ddag}{k_BT}} \left[ \exp\left( \sum_r \frac{s_r}{k_BT} \frac{\delta G}{\delta c_r}\right) - \exp\left( \sum_p \frac{s_p}{k_BT} \frac{\delta G}{\delta c_p} \right)\right]  }   \label{eq:Rphase}
\end{equation}
for the general reaction, Eq. \ref{eq:genreact}, in a  concentrated solution. This is the fundamental expression of thermally activated reaction kinetics that is consistent with nonequilibrium thermodynamics. The reaction rate is a nonlinear function of the thermodynamic driving force,
\begin{equation}
\Delta\mu = \sum_p s_p \frac{\delta G}{\delta c_p} - \sum_r s_r \frac{\delta G}{\delta c_r}.
\end{equation}
This is the most general, variational definition of the free energy of reaction.
For $|\Delta\mu|\ll k_BT$, the rate expression (\ref{eq:Rphase}) can be linearized as
\begin{equation}
    R\sim -k\Delta\mu, \ \ k=\frac{k_0e^{-\mu_{\ddag}^{ex}/k_BT}}{k_BT},
    \label{eq:Rphaselin}
\end{equation}
but more generally, the forward and backward rates have exponential, Arrhenius dependence on the chemical potential barriers. The variational formulation of chemical kinetics, eq \ref{eq:Rphase}, can be applied to any type of reaction (Figure \ref{fig:CHR}), as we now explain.



\subsection{ Heterogeneous chemistry } At an interface, Eq. \ref{eq:Rphase} provides a new reaction boundary condition~\cite{singh2008,burch2009,bai2011,large_acis} 
\begin{equation}
s_i\, A_r \, \hat{n}\cdot \left(\vec{u} \, c_i - \frac{D_i c_i }{k_BT} \nabla \frac{\delta G}{\delta c_i} \right)
= \pm R\left( \left\{ \frac{\delta G}{\delta c_j} \right\} \right)
\label{eq:CHbc}
\end{equation}
($+$ for reactants, $-$ for products; $A_r=$ reaction site area)
for the Cahn-Hilliard (CH) equation~\cite{kom},
\begin{equation}
\frac{\partial c_i}{\partial t} + \vec{u}\cdot \nabla c_i = \nabla\cdot 
\left( \frac{D_i c_i }{k_BT} \nabla \frac{\delta G}{\delta c_i}  \right),  \label{eq:CH}
\end{equation}
expressing mass conservation for the LIT flux Eq. \ref{eq:flux}  with convection in a mean flow $\vec{u}$.  For thermodynamic consistency, $D_i$ is given by Eq. \ref{eq:D}, which reduces Eq. \ref{eq:CH} to the ``modified" CH equation~\cite{nauman2001} in an ideal mixture~\cite{ferguson2012}. This is the ``Cahn-Hilliard reaction (CHR) model''.

\subsection{ Homogeneous chemistry } For bulk reactions, Eq. \ref{eq:Rphase} provides a new source term for  
the CH equation,
\begin{equation}
\boxed{ 
\frac{\partial c_i}{\partial t} + \vec{u}\cdot \nabla c_i = \nabla\cdot 
\left(\frac{D_i c_i }{k_BT} \nabla \frac{\delta G}{\delta c_i}  \right)
\mp \frac{c_s}{s_i} R\left( \left\{ \frac{\delta G}{\delta c_j} \right\} \right) }
\label{eq:CHhomo}
\end{equation}
($c_s=$ reaction sites/volume). The Allen-Cahn equation~\cite{kom} (AC) corresponds to the special case of an immobile reactant ($D_i=0$, $\vec{u} = 0$) evolving according to linear kinetics, eq \ref{eq:Rphaselin}, although the exchange-rate prefactor $k$ is usually taken to be a constant, in contrast to our nonlinear theory. Eq. \ref{eq:CHhomo} is the fundamental equation of non-equilibrium chemical thermodynamics. It unifies and extends the CH and AC equations via a consistent set of reaction-diffusion equations based on variational principles.  Eq. \ref{eq:CHbc} is its integrated form for a reaction localized on a boundary.

\subsection{ Phase transformations}   In the case of immobile reactants with two or more stable states, our general reaction-diffusion equation, eq \ref{eq:CHhomo}, also describes thermally activated phase transformations. The immobile reactant concentrations acts as a non-conserved order parameter, or phase field, representing different thermodynamic states of the same molecular substance. For example, if $\bar{g}(c)$ has two local equilibrium states, $c_A$ and $c_B$, then 
\begin{equation}
\xi=\frac{c-c_A}{c_A-c_B}
\end{equation}
is a phase field with minima at $\xi=0$ and $\xi=1$ satisfying
\begin{equation}
\frac{\partial \xi}{\partial t}  = R\left(\frac{\delta G}{\delta \xi} \right)  \label{eq:genAC}
\end{equation}
This is the ``Allen-Cahn reaction (ACR) model", which is a nonlinear generalization of the AC equation for chemical kinetics~\cite{10.626,singh2008,bai2011,cogswell2012}.

\subsection{ Example:  Adsorption with Condensation }
\begin{figure*}[t]
\includegraphics[width=6in]{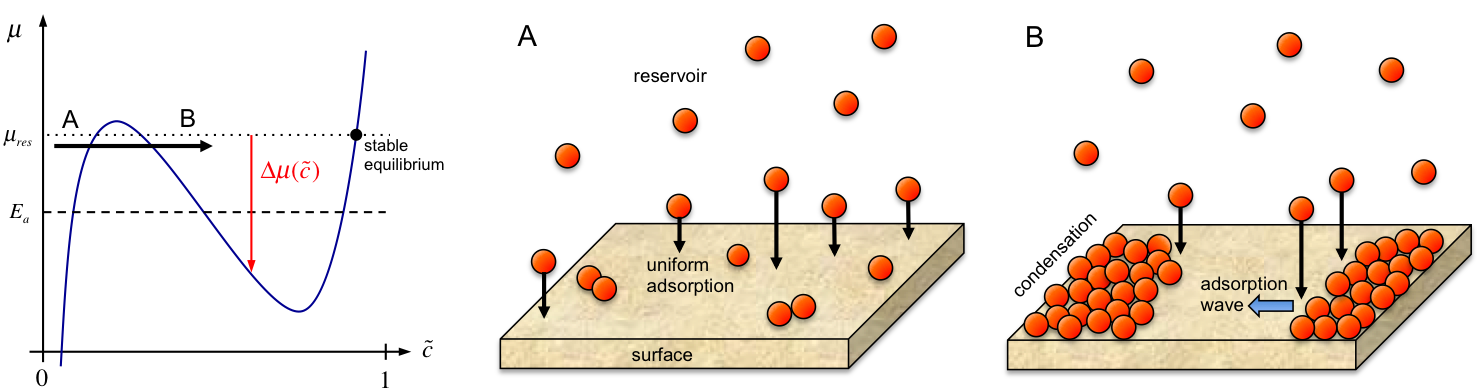}
\caption{ Surface adsorption with condensation when an empty surface is brought into contact with a reservoir ($\mu_{res}=\mu_1=k_BT\ln a > E_{ads} = -k_BT \ln K^\Theta$).  Left: Homogeneous chemical potential of the adsorbed species $\mu$. Right: (A) early-stage uniform adsorption and (B) late-stage adsorption waves nucleated at edges, where the reaction is focused on advancing boundaries of the condensed phase.  \label{fig:condensation}    }
\end{figure*}

To illustrate the theory, we revisit surface adsorption with attractive forces, strong enough to drive adatom condensation (separation into high- and low-density phases) on the surface~\cite{10.626}. Applications may include water adsorption in concrete~\cite{sorption-II} or colloidal deposition in electrophoretic displays~\cite{murau1978}. Following Cahn and Hilliard~\cite{cahn1958}, the simplest model is a regular solution of adatoms and vacancies with pair interaction energy $\Omega$,
\begin{eqnarray}
g = c_s\left\{ k_BT \left[ \tilde{c}\ln\tilde{c} + (1-\tilde{c})\ln(1-\tilde{c})\right] \right. \nonumber \\
\left. + \Omega\tilde{c}(1-\tilde{c})+E_{ads}\tilde{c}\right\}
+ \frac{\kappa}{2}|\nabla\tilde{c}|^2     \label{eq:gregsol} 
\end{eqnarray}
\begin{equation}
\mu = 
k_BT \ln \frac{\tilde{c}}{1-\tilde{c}} + \Omega(1-2\tilde{c}) + E_{ads} - \frac{\kappa}{c_s} \nabla^2 \tilde{c}  \label{eq:muregsol}
\end{equation}
Below the critical point, $T < T_c = \frac{\Omega}{2k_B}$, the enthalpy of adatom attraction (third term, favoring phase separation $\tilde{c}=0,1$) dominates the configurational entropy of adatoms and vacancies (first two terms, favoring mixing $\tilde{c}=\frac{1}{2}$). The gradient term controls spinodal decomposition and stabilizes phase boundaries of thickness $\lambda_b=\sqrt{\frac{\kappa}{c_s\Omega}}$ and interphasial tension $\gamma_b=\sqrt{\kappa c_s \Omega}$. Using Eq. \ref{eq:gamlang} to model the transition state with one excluded site, $s=1$, the ACR model, eq \ref{eq:genAC}, takes the dimensionless form,
\begin{equation}
\frac{\partial \tilde{c}}{\partial \tilde{t}} =  K^\Theta a (1-\tilde{c}) - \tilde{c}\, \exp\left(\tilde{\Omega} (1-2\tilde{c}) 
- \tilde{\kappa} \tilde{\nabla}^2 \tilde{c} \right)
\end{equation}
where $\tilde{t} = k_1 t$, $\tilde{\Omega}=\frac{\Omega}{k_BT}=\frac{2T_c}{T}$, $\tilde{\kappa} = \frac{\kappa}{L^2 c_s k_BT}$ and $\tilde{\nabla} = L\nabla$ (with length scale $L$). 
This nonlinear PDE describes phase separation coupled to adsorption at an interface (Fig. \ref{fig:condensation}), controlled by the reservoir activity $a$.  It resembles a reaction-diffusion equation, but there is no diffusion; instead, $-\tilde{\kappa} \tilde{\nabla}^2 \tilde{c}$ is a gradient correction to the chemical potential, which nonlinearly affects the adsorption reaction rate.   With modifications for charge transfer and coherency strain, a similar PDE describes ion intercalation in a solid host, driven by an applied voltage.

\section{Nonequilibrium Thermodynamics of Electrochemical Systems}

\subsection{ Background }  We thus return to our original motivation -- phase separation in Li-ion batteries (Fig. \ref{fig:wave}). Three important papers in 2004 set the stage:  ~\citet{garcia2004} formulated variational principles for electromagnetically active systems, which unify the CH equation with Maxwell's equations;  \citet{guyer2004a} represented the metal/electrolyte interface with a continuous phase field $\xi$ evolving by AC kinetics~\cite{guyer2004b}; \citet{han2004} used the CH equation to model diffusion in LFP, leading directly to this work.

When the time is ripe for a new idea, a number of scientists naturally think along similar lines.  As described in the Introduction, my group first reported phase-field kinetics (CHR and ACR)~\cite{singh2007,singh2008} and modified Poisson-Nernst-Planck (PNP) equations~\cite{kilic2007b} in 2007, the generalized BV equation~\cite{large_acis} in 2009, and the complete theory~\cite{10.626,bai2011} in 2011. Independently, 
Lai and Ciucci also applied non-equilibrium thermodynamics to electrochemical transport~\cite{lai2010}, but did not develop a variational formulation. They proceeded to generalize BV kinetics~\cite{lai2011a,lai2011b} (citing \citet{singh2008}) but used $\mu$ in place of $\mu^{ex}$ and neglected $\gamma_\ddag^{ex}$.   \citet{tang2011} were the first to apply CHR to ion intercalation with coherency strain, but, like ~\citet{guyer2004b}, they assumed linear AC kinetics.    Recently, Liang et al.~\cite{liang2012} published the BV-ACR equation, claiming that ``in contrast to all existing phase-field models, the rate of temporal phase-field evolution... is considered nonlinear with respect to
the thermodynamic driving force".  They cited my work ~\cite{singh2008,burch2009,bai2011,cogswell2012}   as a ``boundary condition for a fixed electrode-electrolyte interface" (CHR) but overlooked the same BV-ACR equation for the depth-averaged ion concentration~\cite{singh2008,bai2011}, identified as a phase field for an open system~\cite{bai2011,cogswell2012}. They also set $I_0=$constant, which contradicts chemical kinetics (see below).

\subsection{Variational Electrochemical Kinetics}    
We now apply phase-field kinetics to charged species.
The  Gibbs  free energy of ionic materials can be modeled as~\cite{garcia2004,guyer2004a,bazant2011,singh2008,burch2009,bai2011,cogswell2012,samin2011}:
\begin{equation}
G  = G_{mix} + G_{elec} + G_{surf}  + \sum_i \mu_i^{\Theta} \int_V c_i \, dV \label{eq:Gphase} 
\end{equation}
\begin{equation}
G_{mix}  = \int_V  f(\vec{c}) dV + G_{grad}      \label{eq:Gchem}
\end{equation}
\begin{equation}
G_{grad} = \frac{1}{2} \int_V \left( \nabla\vec{\tilde{c}}\cdot \kappa \nabla\vec{\tilde{c}}
-  \nabla\phi \cdot \varepsilon_p \nabla \phi
+  \sigma:\epsilon \right)dV  
\end{equation}
\begin{equation}
G_{elec} =   \int_V \rho_e \phi  dV   +  \oint_A  q_s \phi \, dA
\end{equation}
where $G_{grad}$ is the free energy associated with all gradients; $G_{elec}$ is the energy of charges in the electrostatic potential of mean force, $\phi$; $\vec{c}$ is the set of concentrations (including electrons for mixed ion/electron conductors); $f$ is  the homogeneous Helmholtz free energy density, $\rho_e$ and $q_s$  are the bulk and surface charge densities; $\varepsilon_p$ is the permittivity tensor; and $\sigma$ and $\epsilon$ are the stress and strain tensors. The potential $\phi$ acts as a Lagrange multiplier constraining the total ion densities~\cite{garcia2004,cogswell2012} while enforcing Maxwell's equations for a linear dielectric material ($\frac{\delta G}{\delta \phi}=0$), 
\begin{equation}
 -\nabla \cdot \varepsilon_p \nabla\phi = \rho_e = \sum_i z_i e c_i  \label{eq:Poisson}  
 \end{equation}
\begin{equation}
- \hat{n}\cdot\varepsilon_p \nabla\phi = q_s 
\end{equation}
The permittivity can be a linear operator, e.g.  $\varepsilon_p = \varepsilon_0(1-\ell_c^2\nabla^2)$, to account for electrostatic correlations in ionic liquids~\cite{bazant2011} and concentrated electrolytes~\cite{large_acis,storey2012} (as first derived for counterion plasmas~\cite{santangelo2006,hatlo2010}).
Modified PNP equations~\cite{kilic2007b,large_acis,lai2010,lai2011a} correspond to Eq. \ref{eq:CH} and Eq. \ref{eq:Poisson}.

For elastic solids, the stress is given by Hooke's law, $\sigma_{ij} = C_{ijkl} \epsilon_{kl}$, where $C$ is the elastic constant tensor.  The coherency strain,
\begin{equation}
 \epsilon_{ij}=\frac{1}{2}\left(\frac{\partial u_i}{\partial x_j}  + \frac{\partial u_j}{\partial x_i}\right)-\sum_m \epsilon_{ijm}^0 \tilde{c}_m   \label{eq:strain}
\end{equation}
is the total strain due to compositional inhomogeneity (first term) relative to the stress-free inelastic strain (second term), which contributes to $G_{mix}$. In a mean-field approximation (Vegard's law), each molecule of species $m$ exerts an independent strain $\epsilon_m^0$ (lattice misfit between $\tilde{c}_m=0,1$ with $c_m^\Theta=c_s$).  Since elastic relaxation (sound) is faster than diffusion and kinetics, we assume mechanical equilibrium, $\frac{\delta G}{\delta \vec{u}}=\nabla\cdot\sigma=0$.

For Faradaic reactions Eq. \ref{eq:Faradaic}, the overpotential is the thermodynamic driving force for charge transfer,
\begin{equation}
\boxed{ 
ne \eta = \sum_j s_{j,R} \frac{\delta G}{\delta c_{j,R}}
- \sum_i s_{i,O} \frac{\delta G}{\delta c_{i,O}}-n \frac{\delta G}{\delta c_e},}   \label{eq:etaphase}
\end{equation}
determined by the  electrochemical potentials $\mu_i   = \frac{\delta G}{\delta c_i}$.  For thermodynamic consistency, the diffusivities Eq. \ref{eq:D}, Nernst voltage Eq. \ref{eq:Nernst} and exchange current Eq. \ref{eq:I0} must depend on $\vec{c}$, $\nabla \vec{c}$, and $\sigma$ via the variational activities Eq. \ref{eq:actphase}, given by
\begin{equation}
k_BT\ln a_i = \frac{\partial f}{\partial c_i} - \frac{ \nabla\cdot \kappa \nabla \tilde{c}_i + \sigma : \epsilon^0_i }{c_s} - \nabla \phi \cdot \frac{\partial \varepsilon_p}{\partial c_i}\nabla \phi    \label{eq:ionact}
\end{equation}
for the ionic model above.
The Faradaic current density is 
\begin{equation}
I = I_0 \, F\left( \frac{ ne\eta}{k_BT} \right)   \label{eq:Iphase}
\end{equation}
where 
\begin{equation}
F(\tilde{\eta}) = \left\{ \begin{array}{ll}
e^{-\alpha \tilde{\eta}} - e^{ (1-\alpha) \tilde{\eta}} & \mbox{ Butler-Volmer }  \\
e^{-\tilde{\eta}^2/4\tilde{\lambda}}\left( e^{-\alpha \tilde{\eta}} - e^{(1-\alpha) \tilde{\eta}} \right)   &
\mbox{ Marcus }
\end{array}\right.
\end{equation}
and $I_0$ is given by either Eq. \ref{eq:I0} or Eqs. \ref{eq:marcI0}-\ref{eq:gamgen}, respectively ($\tilde{\lambda}=\frac{\lambda}{k_B T}$) . The charge-transfer rate, $R = \frac{I}{ne}$, defines the CHR and ACR models, Eqs. \ref{eq:CHbc}-\ref{eq:genAC}, for  electrochemical systems.

\subsection{ Example: Metal Electrodeposition }
In models of electrodeposition~\cite{guyer2004a,guyer2004b} and electrokinetics~\cite{gregersen2009}, the solid/electrolyte interface is represented by a continuous phase field $\xi$ for numerical convenience (to avoid tracking a sharp interface).  If the phase field evolves by reactions, however, it has physical significance, as a chemical concentration. For example, consider electrodeposition, $\mbox{M}^{n+} + ne^- \to \mbox{M}$, of solid metal M from a binary electrolyte $\mbox{M}^+ \mbox{A}^-$ with dimensionless concentrations, $\xi=\tilde{c}=c/c_s$ and $\tilde{c}_\pm/c_0$, respectively. 
In order to separate the metal from the electrolyte, we postulate  
\begin{equation}
f = W \left[ h(\tilde{c}) + \tilde{c}(\tilde{c}_+ +\tilde{c}_-)\right] + f_{ion}(\tilde{c}_+,\tilde{c}_-)    
\end{equation}
with $W\gg k_BT$,  where $h=\tilde{c}^2(1-\tilde{c})^2$ is an arbitrary double-welled potential. For a dilute electrolyte, $f_{ion} = k_BT(\tilde{c}_+\ln\tilde{c}_+ + \tilde{c}_-\ln\tilde{c}_-)$, without phase separation~\cite{samin2011}, we include gradient energy only for the metal. The activities Eq. \ref{eq:ionact} for reduced metal
\begin{equation}
c_s k_BT \ln a = W\left[ h^\prime(\tilde{c}) + \tilde{c}_+ + \tilde{c}_-\right] - \kappa \nabla^2 \tilde{c}
- \frac{\partial \varepsilon_p}{\partial \tilde{c}} |\nabla \phi|^2   
\end{equation}
and metal cations 
\begin{equation}
c_0 k_BT \ln a_+ = W \tilde{c} + k_BT \ln \tilde{c}_+ - \frac{\partial \varepsilon_p}{\partial \tilde{c}_+} |\nabla \phi|^2
\end{equation}
define the current density Eq. \ref{eq:Iphase} via 
\begin{equation}
I_0 = K_0 a^\alpha (a_+a_e^n)^{1-\alpha}, \ \ K_0 = \frac{ne k_0 c_s}{\gamma_\ddag}  \label{eq:I0dep}
\end{equation}
\begin{equation}
\eta = \frac{k_BT}{ne} \ln \frac{a}{a_+ a_e^n} - E^\Theta   \label{eq:etametal}
\end{equation}
Note that the local potential for electrons and ions is unique ($\phi=\phi_e$, $\Delta\phi=0$), but integration across the diffuse interface yields the appropriate interfacial voltage.

The ACR equation Eq. \ref{eq:genAC} for $\xi=\tilde{c}$ with Eqs. \ref{eq:Iphase}- \ref{eq:etametal} differs from prior phase-field models~\cite{guyer2004b,liang2012}.  Eq. \ref{eq:I0dep} has the thermodynamically consistent dependence on reactant activities (rather than $I_0=$constant). Coupled with Eq. \ref{eq:CHhomo} for $\tilde{c}_\pm$,  our theory also describes Frumkin corrections to BV kinetics~\cite{biesheuvel2009galvanic,biesheuvel2011} and electro-osmotic flows~\cite{gregersen2009} associated with diffuse charge in the electrolyte.

\subsection{ Example: Ion Intercalation }   
Hereafter, we neglect double layers and focus on solid thermodynamics. Consider cation intercalation, A$^{n+}  + $B$ + n e^- \to $AB, from an electrolyte reservoir ($a_O=$constant) into a conducting solid B ($a_e=$constant) as a neutral polaron ($c_R=c(x,t)$, $z_R=0$). The overpotential Eq. \ref{eq:etaphase} takes the simple form 
\begin{equation}
ne \eta = \frac{ \delta G}{\delta c} - (\mu_O + n\mu_e) = \frac{\delta G_{mix}}{\delta c} + ne \Delta \Phi
\end{equation}
where
\begin{equation}
\Delta \Phi = \Delta\phi - E^\Theta - \frac{k_BT}{ne} \ln a_Oa_e^n
\end{equation}
is the interfacial voltage relative to the ionic standard state. The equilibrium voltage is 
\begin{equation}
ne \Delta \Phi_{eq} = - k_BT \ln a = - \frac{\delta G_{mix}}{\delta c}.
\end{equation}
Note that potentials can be shifted for convenience:  \citet{bai2011} and \citet{ferguson2012} set $\mu^\Theta=0$ for ions, so $\mu=k_BT \ln a=\frac{\delta G_{mix}}{\delta c}$; \citet{cogswell2012} defined ``$\Delta\phi$"$=\Delta \Phi$ and shifted $g$ by $-c \Delta \Phi$, so $e \eta = \frac{\delta G}{\delta c}$. 

Our surface adsorption model Eq. \ref{eq:gregsol} can be adapted  for ion intercalation by setting $E_a = e \Delta\Phi$.
If the transition state  excludes $s$ sites (where $s>1$ could account for the A$^{n+}$ solvation shell) and has strain $-\epsilon_\ddag$, then its activity coefficient Eq. \ref{eq:gamgen} is 
\begin{equation}
\gamma_\ddag = (1-\tilde{c})^{-s} e^{- \tilde{\sigma}:\epsilon_\ddag + \tilde{\lambda}_0/4 }
 \label{eq:gamins}
\end{equation}
where $\tilde{\lambda}_0 = \frac{\lambda_0}{k_BT}$ and  $\tilde{\sigma}=\frac{\sigma}{c_sk_BT}$.
The exchange current Eq. \ref{eq:I0} is 
\begin{equation}
I_0 = n e k(\tilde{c}) \tilde{c}^\alpha(1-\tilde{c})^{s-\alpha}\, e^{ \tilde{\sigma}:\Delta\varepsilon + \alpha \tilde{\Omega}(1-2\tilde{c}) - \alpha  \tilde{\nabla}\cdot\tilde{\kappa}\tilde{\nabla} \tilde{c} } \label{eq:I0int} 
\end{equation}
\begin{equation}
k(\tilde{c}) = k_0 c_s (a_+ a_e^n(\tilde{c}))^{1-\alpha} \, e^{-\tilde{\lambda}_0/4}
\end{equation}
where $a_+$ is the ionic activity in the electrolyte and $\Delta \varepsilon=\epsilon_\ddag-\alpha\epsilon^0$ is the activation strain~\cite{aziz1991}.  For semiconductors, the electron activity $a_e=e^{\Delta E_f/k_BT}$ depends on $\tilde{c}$, if the intercalated ion shifts the Fermi level by donating an electron to the conduction band, e.g. $\Delta E_f \propto (1 + \beta \tilde{c})^{2/d}$ for free electrons in $d$ dimensions (as in LiWO$_3$ with $d=3$ ~\cite{raistrick1981}).

\section{ Application to Li-ion Battery Electrodes }

\subsection{ Allen-Cahn-Reaction Model }

The three-dimensional CHR model Eqs. \ref{eq:CHbc}-\ref{eq:CH} with current density $I=ne R$ given by Eq. \ref{eq:Iphase} and  Eq. \ref{eq:I0int}  describes ion intercalation in a solid particle from an electrolyte reservoir.  In nanoparticles, solid diffusion times (ms-s) are much shorter than discharge times, so a reaction-limited ACR model is often appropriate.  In the case of LFP nanoparticles, strong crystal anisotropy leads to a two-dimensional ACR model over the active (010) facet by depth averaging over $N_s$ sites in the [010] direction~\cite{singh2008,bai2011}.  For particle sizes below 100nm, the concentration tends to be uniform in [010] due to the fast  diffusion~\cite{morgan2004} (uninhibited by Fe anti-site defects~\cite{malik2010}) and elastically unfavorable phase separation~\cite{cogswell2012}. 

\begin{figure*}[t]
\begin{center}
\includegraphics[width=6.5in]{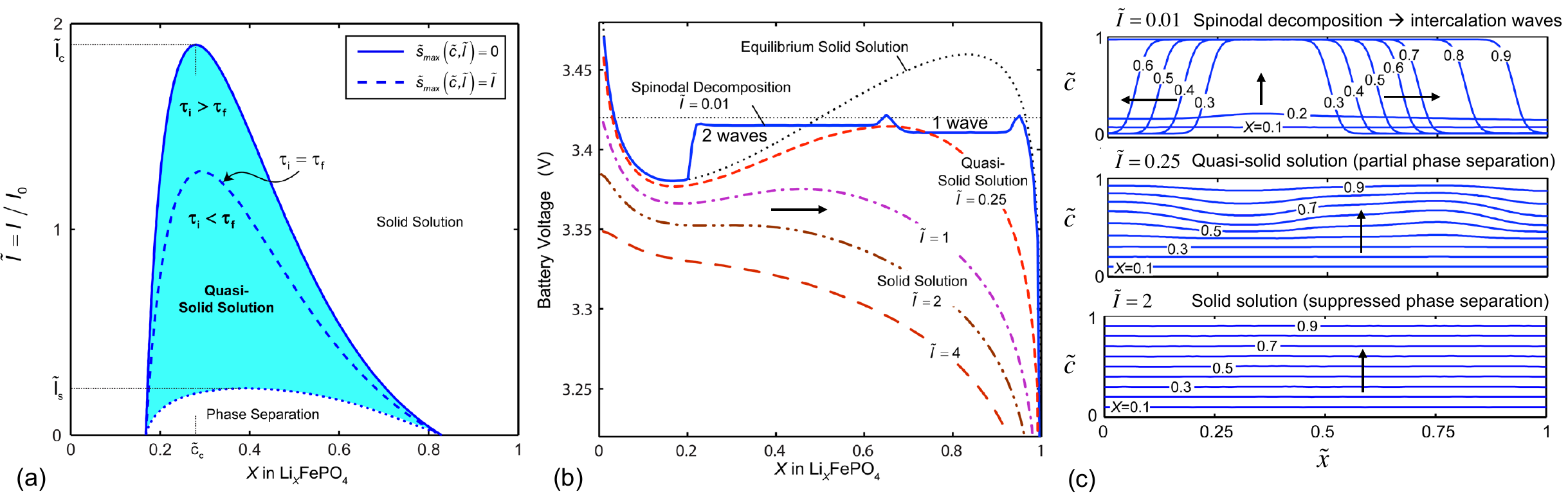}
\caption{ Suppression of phase separation at constant current in an Li-ion battery nanoparticle (ACR model without coherency strain or surface wetting)~\cite{bai2011}. (a) Linear stability diagram for the homogeneous state versus  dimensionless current $\tilde{I} = I/I_0(\tilde{c}=0.5)$  and state of charge $X$.   (b) Battery voltage versus $X$ with increasing $\tilde{I}$. (c) Concentration profiles:  Spinodal decomposition at $\tilde{I}=0.01$ leading to intercalation waves (Fig. \ref{fig:wave}(c)); quasi-solid solution at $\tilde{I}=0.25$; homogeneous filling at $\tilde{I}=2$.  \label{fig:sep} }
\end{center}
\end{figure*}

Using Eq. \ref{eq:Iphase} and Eq. \ref{eq:I0int} with isotropic
$\tilde{\kappa}$, $a_e=$ constant, $\epsilon_\ddag=\alpha\epsilon^0$, $\alpha=\frac{1}{2}$, and $s=1$, the ACR equation, Eq. \ref{eq:genAC}, takes the simple dimensionless form~\cite{bai2011,cogswell2012},
\begin{equation}
\frac{\partial \tilde{c}}{\partial \tilde{t}} = \tilde{I}_0\; F(\tilde{\mu} + \Delta\tilde{\Phi} ) 
\end{equation}
\begin{equation}
\tilde{\mu} = \ln\frac{\tilde{c}}{1-\tilde{c}} +  \tilde{\Omega}(1-2\tilde{c}) - \tilde{\kappa}\tilde{\nabla}^2 \tilde{c} 
- \tilde{\sigma}:\epsilon^0
\end{equation}
\begin{equation}
\tilde{I}_0 = \sqrt{ \tilde{c}  (1-\tilde{c})}\; 
e^{( \tilde{\Omega}(1-2\tilde{c}) - \tilde{\kappa}\tilde{\nabla}^2 \tilde{c})/2 } 
\end{equation}
where  $\Delta\tilde{\Phi}=\frac{ne\Delta\Phi}{k_BT}$, $\tilde{t} = N_s k t$. The total current integrated over the active facet
\begin{equation}
\tilde{I}(\tilde{t}) = \int_{\tilde{A}} \frac{\partial \tilde{c}}{\partial \tilde{t}} \, d\tilde{x}d\tilde{y}
\end{equation}
is either controlled while solving for $\Delta\tilde{\Phi}(\tilde{t})$ (as in Fig. \ref{fig:sep}), or vice versa (where $\tilde{A}$ is the dimensionless surface area of the active facet).
 
\subsection{ Intercalation Waves and Quasi-Solid Solutions }
The theory predicts a rich variety of new intercalation mechanisms. A special case of the CHR model~\cite{singh2008} is  isotropic diffusion-limited intercalation ~\cite{doyle1993,newman_book} with a shrinking-core phase boundary~\cite{srinivasan2004,dargaville2010}, but the reaction-limited ACR model also  predicts intercalation waves (or ``domino cascades"~\cite{delmas2008}), sweeping across the active facet, filling the crystal layer by layer (Fig. \ref{fig:wave}(c)) ~\cite{singh2008,burch2008wave,tang2011,bai2011,cogswell2012}. Intercalation waves result from spinodal decomposition or nucleation at surfaces~\cite{bai2011} and trace out the voltage plateau at low current (Fig. \ref{fig:sep}).

The theory makes other surprising predictions about electrochemically driven phase transformations. \citet{singh2008} showed that intercalation wave solutions of the ACR equation only exist over a finite range of thermodynamic driving force.
Based on bulk free energy calculations, \citet{malik2011} argued for a ``solid solution pathway" without phase separation under applied current, but \citet{bai2011} used the BV ACR model to show that phase separation is suppressed by activation overpotential at high current (Fig. \ref{fig:sep}), due to the reduced area for intercalation on the phase boundary (Fig. \ref{fig:wave}(c)).  Linear stability analysis of homogeneous filling 
predicts a  critical current, of order the exchange current, above which phase separation by spinodal decomposition is impossible. Below this current, the homogeneous state is unstable over a range of concentrations (smaller than the zero-current spinodal gap), but for large currents, the time spent in this region is too small for  complete phase separation. Instead, the particle passes through a transient  ``quasi-solid solution" state, where its voltage and concentration profile resemble those of a homogeneous solid solution. When nucleation is possible (see below), a similar current dependence is  also observed.

\begin{figure*}[t]
\includegraphics[width=6.5in]{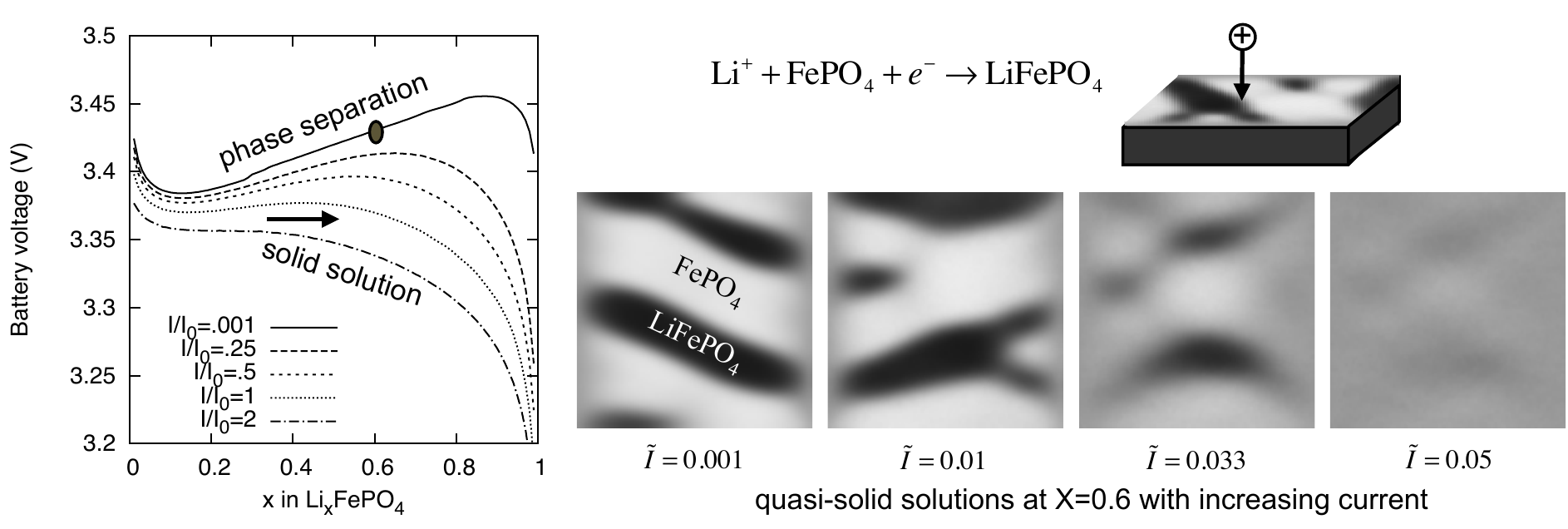}
 \caption{
ACR simulations of galvanostatic discharge in a 100nm Li$_X$FePO$_4$ nanoparticle~\cite{cogswell2012}.  
As the current is increased, transient  quasi-solid solutions (images from the shaded region) transition to homogeneous filling for $\tilde{I}>0.1$, as phase separation is suppressed. 
}
 \label{fig:quasi}
\end{figure*}

For quantitative interpretation of experiments, it is essential to account for the elastic energy~\cite{cogswell2012}.  Coherency strain is a barrier to phase separation (Fig. \ref{fig:quasi}), which tilts the voltage plateau (compared to Fig. \ref{fig:sep}) and reduces the critical current, far below the exchange current. An unexpected prediction is that phase separation rarely occurs {\it in situ} during battery operation in LFP nanoparticles, which helps to explain their high-rate capability and extended lifetime~\cite{bai2011,cogswell2012}.

\begin{figure*}[t]
\includegraphics[width=6.5in]{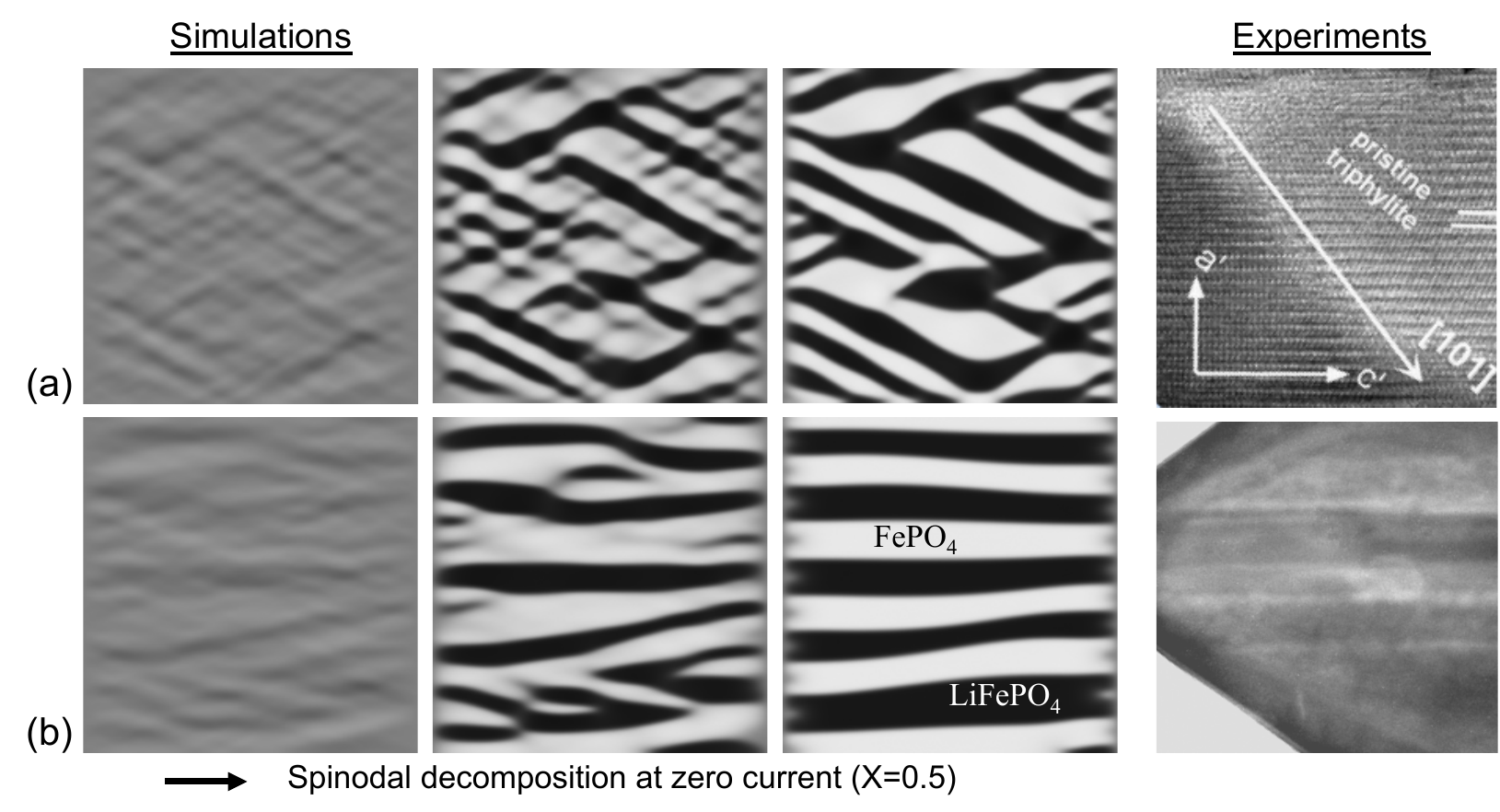}
 \caption{
Phase separation of a 500nm particle of Li$_{0.5}$FePO$_4$ into Li-rich  (black) and Li-poor phases (white) at zero current in ACR simulations~\cite{cogswell2012}, compared with {\it ex situ} experimental images~\cite{ramana2009,chen2006}. (a) Coherent phase separation with [101] interfaces. (b) Semi-coherent phase separation, consistent with observed \{100\} microcracks~\cite{chen2006}.  \label{fig:stripes}
}
\end{figure*}

Phase separation occurs at low currents and can be observed  {\it ex situ} in partially filled particles (Fig. \ref{fig:stripes}).  Crystal anisotropy leads to striped phase patterns in equilibrium~\cite{meethong2007a,vanderven2009,stanton2012}, whose spacing is set by the balance of  elastic energy (favoring short wavelengths at a stress-free boundary) and interfacial energy (favoring long wavelengths to minimize interfacial area)~\cite{cogswell2012}.  \citet{stanton2012} predicted that simultaneous positive and negative eigenvalues of $\epsilon_0$ make phase boundaries  tilt with respect to the crystal axes. In LFP, lithiation causes contraction in the [001] direction and expansion in the [100] and [010] directions~\cite{chen2006}.  Depending on the degree of coherency, \citet{cogswell2012} predicted phase morphologies in excellent agreement with experiments (Fig. \ref{fig:stripes}) and  inferred the gradient penalty $\kappa$ and the LiFePO$_4$/FePO$_4$ interfacial tension (beyond the reach of molecular simulations) from the observed stripe spacing.

\subsection{ Driven Nucleation and Growth }  

The theory can also quantitatively predict nucleation dynamics driven by chemical reactions.  Nucleation is perhaps the least understood phenomenon of thermodynamics. In thermal phase transitions, such as boiling or freezing, the critical nucleus is controlled by random heterogeneities, and its energy is over-estimated by classical spherical-droplet nucleation theory.  Phase-field models address this problem, but often lack sufficient details to be predictive. 

\begin{figure*}[t]
\includegraphics[width=6.5in]{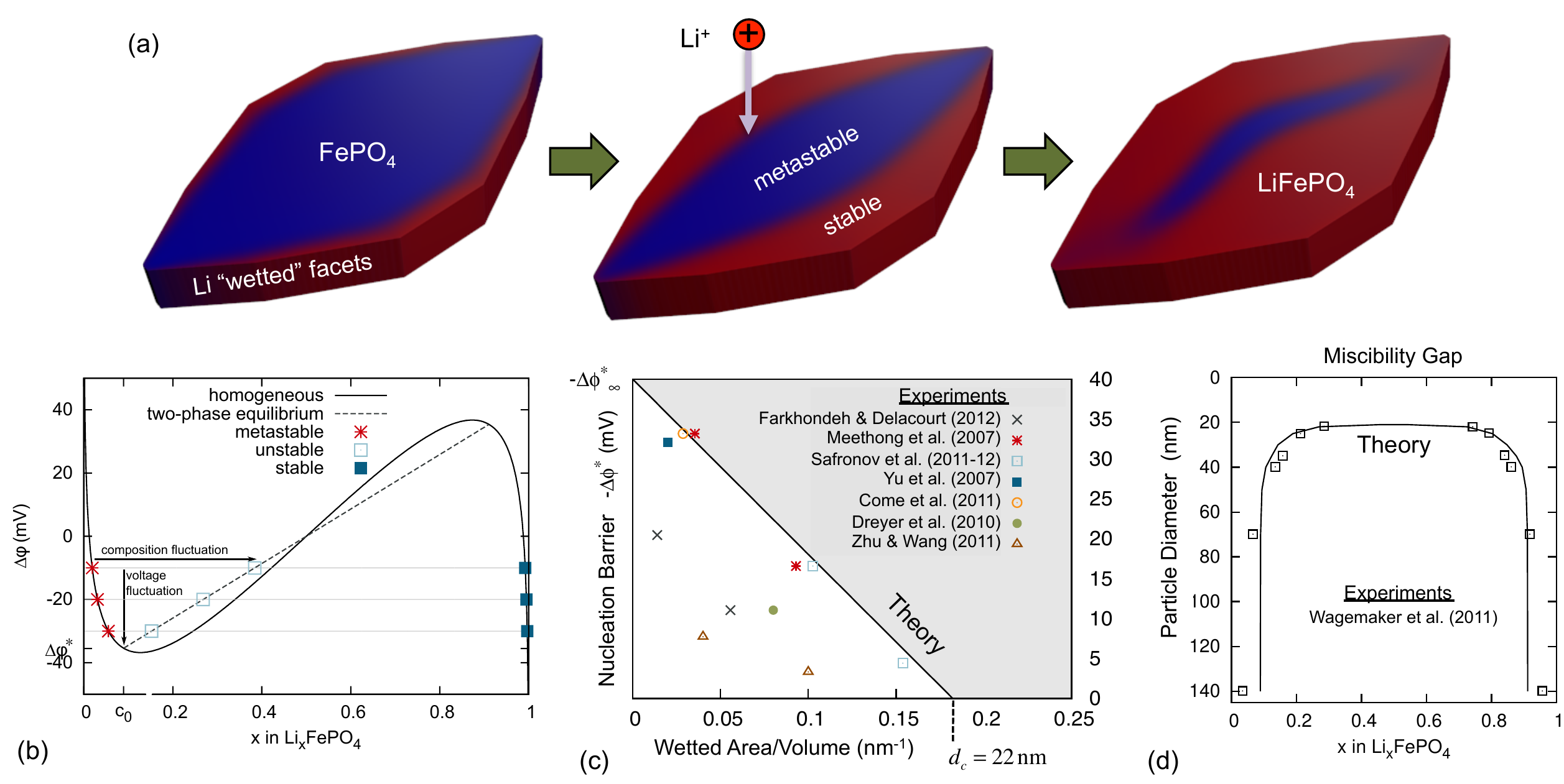}
 \caption{
(a) ACR simulation of galvanostatic nucleation in a realistic LFP nanoparticle shape (C3)~\cite{smith2012} with a 150 nm $\times$ 76 nm top (010) active facet~\cite{cogswell2013}. Surface ``wetting" of the side facets by lithium nucleates intercalation waves that propagate across the particle (while bending from coherency strain) after the voltage exceeds the coherent miscibility limit. (b) Discharge plot indicating nucleation by fluctuations in voltage or composition~\cite{cogswell2013}. (c) Collapse of experimental data for the nucleation voltage by the theory, without any fitting parameters~\cite{cogswell2013}. (d) Size dependence of the miscibility gap, fitted by the theory~\cite{cogswell2012}. }
 \label{fig:nuc}
\end{figure*}

For battery nanoparticles, nucleation turns out to be more tractable, in part because the current and voltage can be more precisely controlled than heat flux and temperature. More importantly, the critical nucleus has a well-defined form, set by the geometry, due to strong surface ``wetting" of crystal facets by different phases. \citet{cogswell2013} showed that nucleation in binary solids occurs at the coherent miscibility limit, as a surface layer becomes unstable and propagates into the bulk. The nucleation barrier, $E_b=- e\Delta\Phi$ is set by coherency strain energy (scaling with volume) in large particles and reduced by surface energy (scaling with area) in nanoparticles. The barrier thus decays with the wetted area-to-volume ratio $A/V$ and vanishes at a critical size, below which nanoparticles remain homogeneous in the phase of lowest surface energy.  

The agreement between theory and experiment -- without fitting any parameters --  is impressive (Fig. \ref{fig:nuc}). Using our prior ACR model~\cite{cogswell2012} augmented only by {\it ab initio} calculated surface energies (in Eq. \ref{eq:wet}), the theory is able to collapse $E_b$ data for LFP versus $A/V$, which lie either on the predicted line or below (e.g. from heterogeneities, lowering $E_b$, or missing the tiniest nanoparticles, lowering $A/V$)~\cite{cogswell2013}. This resolves a major controversy, since the data had seemed inconsistent ($E_b=2.0-37$ mV), and some had argued for~\cite{singh2008,oyama2012,bai2013} and others against the possibility of nucleation (using classical droplet theory)~\cite{malik2011}.  The new theory also predicts that the nucleation barrier (Fig. \ref{fig:nuc}(c)) and miscibility gap (Fig. \ref{fig:nuc}(d)) vanish at the same critical size, $d_c \approx 22$ nm, consistent with separate Li-solubility experiments~\cite{wagemaker2011}.

\subsection{ Mosaic Instability and Porous Electrodes }

These findings have important implications for porous battery electrodes, consisting of many phase separating nanoparticles. The prediction that small particles transform before larger ones is counter-intuitive (since larger particles have more nucleation sites) and opposite to classical nucleation theory. The new theory could be used to predict mean nucleation and growth rates in a simple statistical model ~\cite{bai2013} that fits current transients in LFP~\cite{oyama2012} and guide extensions to account for the particle size distribution. 

\begin{figure*}[t]
\includegraphics[width=6in]{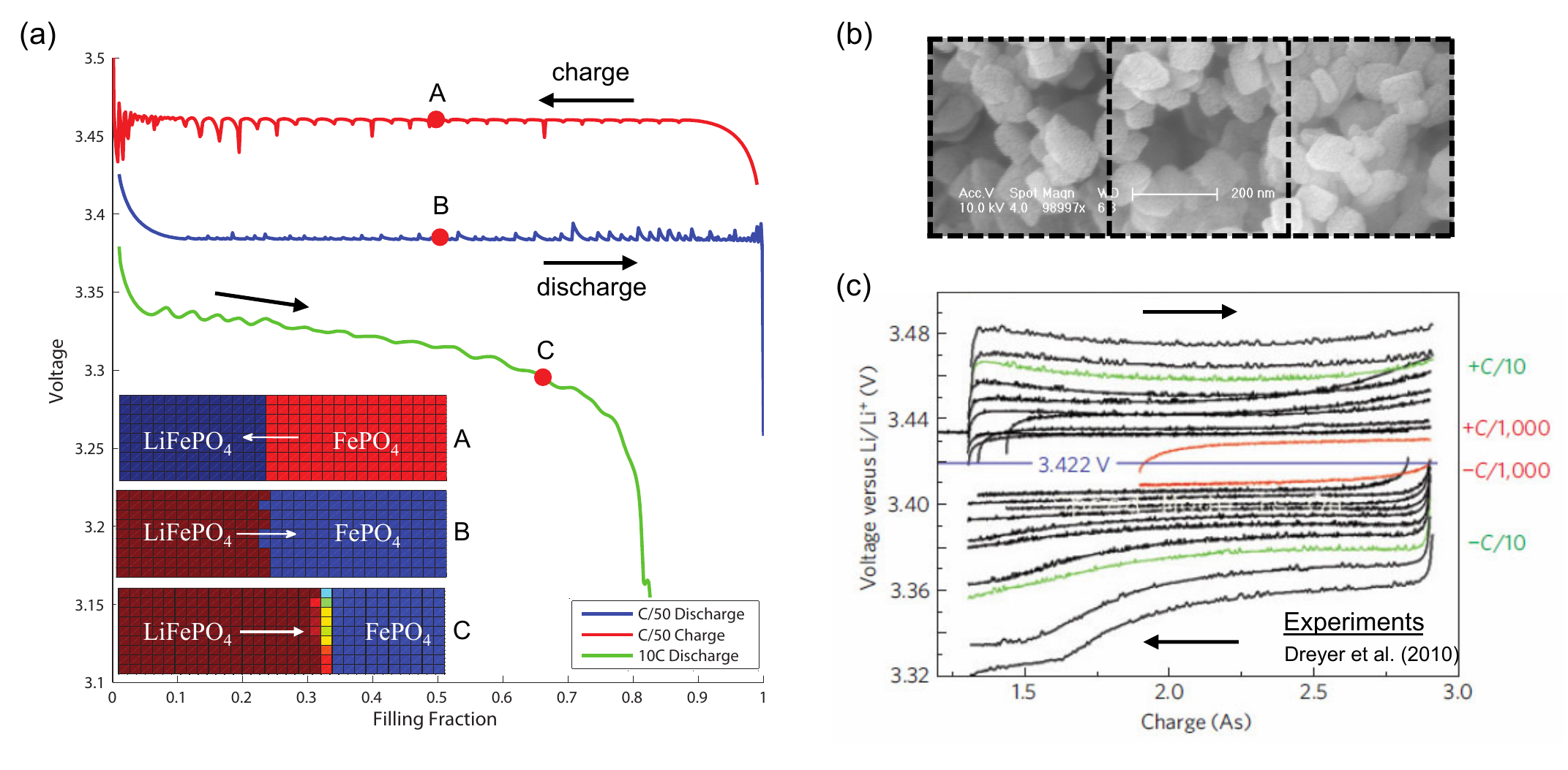}
 \caption{ 
 Finite-volume simulations of a porous LFP cathode (T. Ferguson~\cite{ferguson2012}). (a) Voltage versus state of charge at different rates with profiles of the mean solid Li concentration (A-C), separator on the left, current collector on the right. (b) SEM image of LFP nanoparticles represented by three finite volumes (P. Bai). (c) Experiments revealing a zero-current gap between noisy charge and discharge voltage plateaus (From ~\citet{dreyer2010}).  \label{fig:porous}
}
\end{figure*}

Discrete, random transformations also affect voltage transients. Using the CHR model~\cite{burch2009} for a collection of particles in a reservoir, \citet{burch_thesis} discovered the ``mosaic instability", whereby particles switch from uniform to sequential filling after entering the miscibility gap.  Around the same time, \citet{dreyer2010} published a simple theory of the same effect (neglecting phase separation within particles) supported by experimental observations of voltage gap  between charge/discharge cycles in LFP batteries (Fig. \ref{fig:porous}(c)), as well as pressure hysteresis in ballon array ~\cite{dreyer2011}. 

The key ingredient missing in these models is the transport of ions (in the electrolyte) and electrons (in the conducting matrix), which mediates interactions between nanoparticles and becomes rate limiting at high current.  Conversely, the classical description of porous electrodes, pioneered by Newman~\cite{newman_book,doyle1993}, focuses on transport, but mostly neglects the thermodynamics of the active materials~\cite{lai2011a,ferguson2012}, e.g. fitting~\cite{srinivasan2004}, rather than deriving~\cite{dreyer2010,lai2010,lai2011b,bai2011}, the voltage plateau in LFP.   These approaches are unified by non-equilibrium chemical thermodynamics~\cite{ferguson2012}.
Generalized porous electrode theory is constructed by formally volume averaging over the microstructure to obtain macroscopic reaction-diffusion equations of the form Eq. \ref{eq:CHhomo} for three overlapping continua -- the electrolyte, conducting matrix, and active material -- each containing a source/sink for Faradaic reactions, integrated over the internal surface of the active particles, described by the CHR or ACR model.  

The simplest case is the ``pseudo-capacitor approximation" of fast solid relaxation (compared to reactions and macroscopic transport), where the active particles remain homogeneous. Using our model for LFP nanoparticles~\cite{cogswell2012}, the porous electrode theory predicts the zero-current voltage gap, without any fitting (Fig. \ref{fig:porous}). (Using the mean particle size, the gap is somewhat too large, but this can be corrected by size-dependent nucleation (Fig. \ref{fig:nuc}), implying that smaller particles were preferentially cycled in the experiments.) Voltage fluctuations at low current correspond to discrete sets of transforming particles. For a narrow particle size distribution, mosaic instability sweeps across the electrode from the separator as a narrow reaction front (Fig. \ref{fig:porous}(a) inset). As the current is increased, the front width grows, and the active material transforms more uniformly across the porous electrode, limited by electrolyte diffusion. A wide particle size distribution also broadens the reaction front, as particles transform in order of increasing size. These examples illustrate the complexity of phase transformations in porous media driven by chemical reactions. 

\section{ Conclusion }

This Account describes a journey along the ``middle way"~\cite{laughlin2000}, searching for organizing principles of the mesoscopic domain between individual atoms and bulk materials.  The motivation to understand phase behavior in Li-ion battery nanoparticles gradually led to a theory of collective kinetics at length and time scales in the ``middle", beyond the reach of both molecular simulations and macroscopic continuum models.  The work leveraged advances in {\it ab initio} quantum-mechanical calculations and nanoscale imaging, but also required some new theoretical ideas. 

Besides telling the story, this Account synthesizes my work as a general theory of chemical physics,  which transcends its origins in electrochemistry. The main result, Eq. \ref{eq:CHhomo}, generalizes the Cahn-Hilliard and Allen-Cahn equations for reaction-diffusion phenomena. The reaction rate  is a nonlinear function of the species activities and the free energy of reaction (Eq. \ref{eq:Rgen}) via variational derivatives of the Gibbs free energy functional (Eq. \ref{eq:Rphase}), which are consistently defined for non-equilibrium states, e.g. during a phase separation. For charged species, the theory generalizes the Poisson-Nernst-Planck equations of ion transport, the Butler-Volmer equation of electrochemical kinetics (Eq. \ref{eq:I0}), and the Marcus theory of charge transfer (Eq. \ref{eq:marcus}) for concentrated electrolytes and ionic solids.  

As its first application, the theory has predicted new intercalation mechanisms in phase-separating battery materials, exemplified by LFP:
\begin{itemize}
\item  intercalation waves in anisotropic nanoparticles at low currents (Fig. \ref{fig:sep}); 
\item quasi-solid solutions  and suppressed phase separation at high currents (Fig. \ref{fig:quasi});
\item  relaxation to striped phases in partially filled particles (Fig. \ref{fig:stripes});
\item size-dependent nucleation by surface wetting (Fig. \ref{fig:nuc});
\item mosaic instabilities and reaction fronts in porous electrodes (Fig. \ref{fig:porous});
\end{itemize}
These results have some unexpected implications, e.g. that battery performance may be improved with elevated currents and temperatures, wider particle size distributions, and coatings to alter surface energies.
The model successfully describes phase behavior of LFP cathodes, and my group is extending it to graphite  anodes (``staging" of Li intercalation with $\geq 3$ stable phases) and air cathodes (electrochemical growth of Li$_2$O$_2$).

The general theory may find many other applications in chemistry and biology. For example, the adsorption model (Fig. \ref{fig:condensation}) could be adapted for the deposition of charged colloids on transparent electrodes in electrophoretic displays.  The porous electrode model (Fig. \ref{fig:porous}) could be adapted for sorption/desorption kinetics in nanoporous solids, e.g. for drying cycles of cementitious materials, release of shale gas by hydraulic fracturing, carbon sequestration in zeolites, or ion adsorption and impulse propagation in biological cells.  The common theme is the coupling of chemical kinetics with non-equilibrium thermodynamics.

\section*{Acknolwedgements}
This work was supported by the National Science Foundation under Contracts DMS-0842504 and DMS-0948071 and by the MIT Energy Initiative and would not have been possible without my postdocs (D. A. Cogswell, G. Singh)  and students (P. Bai, D. Burch, T. R. Ferguson, E. Khoo, R. Smith, Y. Zeng). P. Bai noted the Nernst factor in Eq. \ref{eq:alpha}.

\noindent
\section*{Bibliographical Information}
{\bf Martin Z. Bazant} received his B.S. (Physics, Mathematics, 1992) and M.S. (Applied Mathematics, 1993) from the University of Arizona and Ph.D. (Physics, 1997) from Harvard University.  He joined the faculty at MIT in Mathematics in 1998 and Chemical Engineering in 2008. His honors include an Early Career Award from the Department of Energy (2003), Brilliant Ten from {\it Popular Science} (2007), and Paris Sciences Chair (2002,2007) and Joliot Chair (2008,2012) from ESPCI (Paris, France).

\bibliography{elec37}

\end{document}